\documentclass[preprintnumbers,showpacs,superscriptaddress,amsmath,amssymb,nofootinbib]{revtex4-1}
\usepackage{amsmath}
\usepackage{mathbbol}
\usepackage[pdftex]{graphicx}
\usepackage{dcolumn}
\usepackage{bm}
\usepackage{SIunits}
\usepackage{color}

\providecommand \citenamefont [1]{#1}%

\begin{document}

\title{Superconductivity with excitons and polaritons}
\author{F.~P.~Laussy}
\altaffiliation{Presently at Walter Schottky Institut, Technische Universit\"at M\"unchen, Am Coulombwall, 3 D-85748, Garching, Germany.}
\affiliation{School of Physics and Astronomy, University of Southampton, Highfield, Southampton, SO17 1BJ, UK.}
\author{T.~Taylor}
\affiliation{School of Physics and Astronomy, University of Southampton, Highfield, Southampton, SO17 1BJ, UK.}
\author{I.~A.~Shelykh}
\affiliation{Science Institute, University of Iceland, Dunhagi-3, IS-107, Reykjavik, Iceland.}
\affiliation{International Institute for Physics, Av. Odilon Gomes de Lima, 1722, CEP: 59078-400, Capim Macio, Natal RN Brazil.}
\author{A.~V.~Kavokin}
\affiliation{School of Physics and Astronomy, University of Southampton, Highfield, Southampton, SO17 1BJ, UK.}

\begin{abstract}
  A Bose--Einstein condensate of exciton polaritons coexisting with a
  Fermi gas of electrons has been recently proposed as a promising
  system for realisation of room-temperature superconductivity
  [Phys.~Rev.~Lett., \textbf{104}, 106402 (2010)]. In order to find
  the optimum conditions for exciton and exciton-polariton mediated
  superconductivity, we study the attractive mechanism between
  electrons of a Cooper pair mediated by the exciton and
  exciton-polariton condensate and analyze the gap equation that
  follows. We specifically address microcavities with embedded
  $n$-doped quantum wells as well as coupled quantum wells hosting a
  condensate of spatially indirect excitons, put in contact with a
  two-dimensional electron gas. We show that engineering of the
  interaction in these peculiar Bose-Fermi mixtures is complex and
  sometimes counterintuitive, but leaves much freedom for
  optimization, making promising the realization of high-temperature
  superconductivity in multilayer semiconductor structures.
\end{abstract}

\maketitle

\affiliation{School of Physics and Astronomy, University of Southampton, Southampton,
SO171BJ, United Kingtom}

\affiliation{School of Physics and Astronomy, University of Southampton, Southampton,
SO171BJ, United Kingtom}

\affiliation{Science Institute, University of Iceland, Dunhagi 3, IS-107,
Reykjavik, Iceland}

\affiliation{School of Physics and Astronomy, University of Southampton, Southampton,
SO171BJ, United Kingtom}

\section{Introduction}

The electron gas undergoes, in some conditions, a phase transition to bound
pairs of electrons (the so-called \emph{Cooper} pairs), that replace
electrons as the fundamental agent of the electronic properties. The Cooper
pairs are, from the point of view of their electric charge, objects
qualitatively identical to the underlying electrons. From the point of view
of their spin, on the other hand, they become integer-spin particles, that
is, from the spin-statistics theorem, bosons rather than
fermions. This shift of statistical paradigm of the carriers, from Fermi to
Bose-statistics, results in the outstanding behaviour of superconductivity,
that is, conduction of electric charge by a macroscopic coherent
wavefunction (akin to a Bose-Einstein condensate). It has taken some time to
capture the fundamental and universal features of this phenomenon and set
them apart from particularities proper to certain cases only. The gap of
excitations, responsible for zero resistivity, for instance, results from
the long-range nature of the Coulomb interaction, but gapless 
superconductivity is also possible. One of the central, fundamental concepts 
of superconductivity is that of a coherent quantum state of charged bosons. 
Although superconductivity was discovered empirically, and its theoretical
construction consisted in assembling a puzzle, it is now possible to
envisage engineering superconducting phases in other systems, based on this
understanding of condensation of charged bosons. If superconducting phases
can be identified in other systems, progresses will be quick for the
understanding of cuprate superconductivity, which still eludes compelling
theoretical explanation of its intrinsic mechanism.

A system that is making rapid and impressive progress in terms of
creating and controlling macroscopic quantum states is that of
microcavity exciton-polaritons~\cite{kavokin_book07a} (see
\cite{deng10a} for a review).  These quasi-particles that combine
properties of light (cavity photons) and matter (quantum well
excitons) have been noted for their predisposition to accumulate in
macroscopic number in a single or few quantum states~\cite%
{imamoglu96a, baumberg00a}. They have many advantages from a practical
point of view, such as their 2D geometry, which allows 
straightforward manipulation by lasers impinging at an angle, and
their short lifetime, which allows continuous monitoring of the
system, reconstructing its internal dynamics also by angle-resolved
spectroscopy~\cite{skolnick98a}. The pumping can be either coherent
(driving states in parametric scattering
configurations)~\cite{savvidis00a,stevenson00a,ciuti01a,ciuti05a,shelykh06a}
or incoherent (with a constant flow of unrelated particles relaxing
into the ground state )~\cite%
{kavokin03a,laussy04c,szymanska06a,kasprzak06a,balili07a,delvalle09c}. In
nitride systems, the formidable claim has been made of room
temperature Bose-Einstein
condensation~\cite{christopoulos07a,baumberg08a}. Recently, there has
been great interest in propagation of polariton fluids~\cite%
{amo09a,sanvitto10a} and their superfluid properties~\cite{amo09b},
with reports of quantized vorticity~\cite{lagoudakis08a,lagoudakis09a}
and persistent currents~\cite{sanvitto10b}.

These rising stars of macrosopic coherence have also been proposed to
service another much sought after quantum phase at high temperature:
superconductivity. Polariton condensates cannot conduct electric
current themselves, being neutral particles. One of the proposed
implementations involves \textquotedblleft quatrons\textquotedblright\
(or quadrions) rather than polaritons~\cite{kavokin07b}. Quatrons are
bound states of two electrons with a polariton. They remain bosons
from spin-addition rules but carry an electric charge. A Bose
condensate of quatrons would, through its superfluid propagation,
exhibit superconductivity. To date, however, the existence of the
quatron, predicted theoretically~\cite{kavokin07b}, has not been found
experimentally. Recently, we have approached the problem from another,
and more conventional, angle, that of the BCS mechanism~\cite%
{laussy10a} with an important new feature: replacement of phonons by
Bose-condensed exciton-polaritons in the role of a binding agent
between electrons. In the present paper, we generalise the model
proposed in \cite%
{laussy10a} to describe a wide range of hybrid Bose-Fermi
semiconductor systems where a Bose-Einstein condensate of neutral
quasiparticles (excitons or exciton-polaritons) coexists with a Fermi
sea of electrons. We show that indeed such systems are promising for
observation of (high-temperature) superconductivity. Moreover, as the
superconducting gap and critical temperature appear to be very
sensitive to the concentration of bosons in the system and the latter
may be controlled by direct optical excitation of exciton, light
induced superconductivity in semiconductor heterostructures appears to
be possible. In this work we closely follow the BCS\ approach,
generalised and adapted to the case of superconductivity
mediated by a Bose-Einstein condensate. Solving the gap equation in
this case turns out to be a non-trivial problem, requiring careful
analysis.

\subsection{BCS with a Bose-Einstein condensate as a binding agent}

BCS is a pillar of superconductivity theory, which relies on three
main tenets:

\begin{enumerate}
  \addtolength{\itemsep}{-.5\baselineskip}
\item Instability of the Fermi sea,
\item Existence of an attractive interaction,
\item Condensation of charged bosons.
\end{enumerate}

These are the three insights that were mainly contributed by Cooper, 
Bardeen and Schrieffer, respectively, and that they could assemble into the BCS
theoretical edifice and exploit to reproduce strikingly or predict
successfully most of the superconductivity phenomenology.

The first point follows from Cooper's observation~\cite{cooper56a} that an
arbitrary small attractive interaction between two electrons on top of the
Fermi sea leads to a bound state (the Cooper pair), thanks to the truncation
of the momentum space for states with wavevector~$k>k_\mathrm{F}$ (Fermi
wavevector). This is a general result, that follows from Bethe-Goldstone
equation for the two-electron problem.

The second point is the identification of an effective attraction between
electrons that normally experience bare Coulomb repulsion. This
attraction is attributed for conventional superconductors to an interaction
through phonons, the Bardeen-Pines potential~\cite{bardeen55b}, that
consists, vividly, of one electron wobbling the lattice at a first time,
which affects another electron at a second time (and at a larger timescale,
since the lattice dynamics is much slower than that of electrons)~\cite%
{bardeen51a}. If the frequency of the lattice vibration is smaller than that
of the propagating electron, the net effect results in an effective
attraction [Leggett offers an insightful toy model of coupled oscillators to
capture the essence of the interaction character (attractive or repulsive)~%
\cite{leggett_book06a}].

The last point is the so-called BCS state, which is a coherent superposition
of paired bound states and which brings the two-particle Cooper effect to a
collective behaviour of all electrons in the system~\cite{bardeen57b}.

With these three ingredients put together, the BCS theory is
complete~\cite%
{bardeen57a}. For our purposes, points~1 and~3 will be regarded as
fundamental and well established features of (BCS type of)
superconductivity. Point~2, that might appear a mere desiderata for
point~1 to apply, leaves us room for identifying and designing new
types of attractive potentials, optimizing the range of applicability
and strength so as to obtain robust superconductivity (e.g.,
holding at high temperatures or high magnetic fields) or its
manifestation in a new class of systems (in microcavities).  Bardeen
himself, with coworkers~\cite{allender73a}, investigated possibilities
to engineer a more robust BCS in a bilayer structure where excitons
replace phonons as mediators of the interaction. The idea of
substituting phonons by excitons was pioneered by
Little~\cite{little64a} and developed by
Ginzburg~\cite{sp_ginzburg70a}, who coined the term and 
theorized the possibility of \emph{high-temperature superconductivity},
much before it came to fruition with cuprates~\cite%
{bednorz86a}. Those, however, exploit another (still unknown)
mechanism different to BCS~\cite{leggett06a}.

In the following, we revisit the Ginzburg mechanism, based on BCS,
with emphasis on maximizing the strength of interaction between
electrons, so as to maintain their binding, and therefore
superconductivity, to higher temperatures. The text is organized as
follows: in Section~\ref%
{sec:TueMay25181652BST2010}, we give a short overview of the BCS
mechanism and the exciton (Ginzburg) mechanism, outlining the points
of special interest in our case. In particular, we introduce the gap
equation. In Section~\ref{sec:TueMay25181751BST2010}, we introduce our
hybrid Bose-Fermi system configuration, its Hamiltonian and
microscopic interactions, and the effective electron-electron
Hamiltonian that results from a mean field approximation for the
condensate and the usual Fr\"{o}lich transformation.  We obtain the
shape of the effective electron-electron interaction $U$, that we find
to be quite different in character to the Cooper (square well)
potential. We also consider possible variations of our scheme, namely,
a microcavity with a condensate of exciton-polaritons and a condensate
of indirect excitons in coupled quantum wells \cite{butov02a}. The
system of coupled quantum wells explored by several groups
\cite{butov04a, rapaport04a} might be easier to realize and study (it
does not need a cavity) and presents some interesting differences as
compared to the polariton system. On the other hand, polaritons
condense at much higher temperatures \cite{baumberg08a}. In
Section~\ref%
{sec:TueMay25182122BST2010}, we study the gap equation for a Bose
condensate mediated effective interaction. Because the potential is
not positive-definite, the problem is not well-posed numerically. We
propose a simplified potential and an approximate solution of the gap
equation, that we motivate by studying its validity on
well-established approximations. We obtain the critical temperature in
this case. In Section~\ref%
{sec:TueMay25182435BST2010}, we give our conclusions and perspective
on this new application of excitons and polaritons and discuss how to
measure the effect experimentally.


\section{BCS and Ginzburg mechanisms}

\label{sec:TueMay25181652BST2010}

Superconductivity is a fundamental property of solids. At low enough
temperatures, most metals superconduct. As the mechanism is rooted in quantum
mechanics, temperatures are expected to be very low and indeed this is the
case for all metals. After considerable theoretical efforts from various
groups, a compelling theoretical model was assembled by Bardeen, Cooper and
Schrieffer, the so-called BCS theory~\cite{bardeen51a}. The model provides
the critical temperature: 
\begin{equation}
k_{\mathrm{B}}T_{\mathrm{C}}=\Theta e^{-\frac{1}{g}}
\label{eq:FriMay28101558BST2010}
\end{equation}%
where, in conventional superconductivity, $\hbar \omega _{\mathrm{D}}$
is the Debye energy, and $g=\mathcal{N}(0)V$ with $\mathcal{N}(0)$ the
density of electrons at the Fermi energy and $V$ the electron-phonon
coupling strength. The Debye energy is, in good approximation, the
maximum energy that can be carried by a phonon. The mechanism
therefore relies heavily on phonons, as was realized empirically
before the advent of BCS (good conductors, for instance, are bad
superconductors, since $V$ is small, or through the isotope effect,
which correlates critical temperature with weight of the crystal
atoms, and thus with resonance frequency). The exponential form and
the presence of $\mathcal{N}(0)$ show that the effect is a collective
one involving all electrons, which have formed a new phase of matter
that cannot be approached perturbatively by the independent electron
pictures (since $f(z)=e^{-1/z}$ has no Taylor expansion around
zero). Based on the theory, and accumulated experience, it was widely
accepted that critical temperatures would not exceed a few tens of
Kelvins, since the Debye energy, which can be quite large in some
systems (hundreds of Kelvins), is exponentially reduced. In all
conventional superconductors $\mathcal{N}(0)V\ll 1$ (the so-called
\emph{weak-coupling} regime).

Ginzburg made the obvious but daring assumption that to achieve higher
critical temperatures---crucial for technical applications which can easily
be understood to be momentous---it is enough to find a system where $\Theta $
and/or $g$ are increased. Replacing phonons by excitons, for instance,
Ginzburg found that values $\Theta \approx 10^{3}$, $10^{4}$K as well as $%
g\approx \frac{1}{5}$, $\frac{1}{3}$ are obtained, yielding temperatures of
several hundred Kelvin~\cite{ginzburg82a}. High critical temperatures have
been later reported in cuprates~\cite{bednorz86a} and nowadays, temperatures
as high as~125K are reported in systems such as Tl-Ba-Cu-oxide. Just as
superconductivity in metals was discovered ahead of theory, cuprate
superconductivity does not appear to follow the BCS pattern~\cite%
{monthoux07a} even with substitution of the mediating field as proposed by
Ginsburg. All attempts to date to realize the exciton mechanism or a
variation of it have remained fruitless, but it is this mechanism, rooted
in BCS, which we consider in this text. Before
we turn to the mechanism we propose, which consists in substituting
the phonons of conventional BCS, or the excitons of Ginzburg scheme,
by a\ Bose-Einstein condensate (BEC) of excitons or
exciton-polaritons, we recall that the starting point of a microscopic
derivation of superconductivity is the gap-equation:
\begin{equation}  \label{eq:TueApr6172010BST2010}
\Delta_\mathbf{k}=-\sum_{\mathbf{k}^{\prime }}U_{\mathbf{k}\mathbf{k}%
^{\prime }}\frac{\Delta_\mathbf{k}^{\prime }}{2E}
\end{equation}
where~$U_{\mathbf{k}\mathbf{k}^{\prime }}$ is the effective
interaction between electrons with wavevectors~$\mathbf{k}$
and~$\mathbf{k}'$ and energy $E$. The gap~$\Delta$ can be identified
as the macroscopic wavefunction of a Cooper pair, which is also the
order parameter for the superconducting phase. If it is nonzero, the
system is in the superconducting state. A realistic microscopic
treatment of $U$ is very complicated. A simplified version is provided
by the Jellium model, which is a toy model of a metal that gives
predominance to electron-electron interactions, that is, in
particular, the underlying crystal is approximated as a uniform
(structureless) background (like a ``jelly'') in which the interacting
electron gas evolves under its own self-interactions and the overall
charge cancellation of the background. A popular effective interaction
is given by the Bardeen-Pines potential~\cite{leggett_book06a} which
is derived from the microscopic form of the electron-lattice
interaction:
\begin{subequations}
\begin{align}  \label{eq:SatOct31152205GMT2009}
U_\mathrm{BP}(\omega,\mathbf{q})&=\frac{\kappa_0}{1+q^2/q_\mathrm{TF}^2}%
\left\{1+\frac{\omega_\mathrm{ph}(q)^2}{\omega^2-\omega_\mathrm{ph}(q)^2}%
\right\}\,, \\
&=\displaystyle\frac{\kappa_0}{(1+\frac{q^2}{\kappa_\mathrm{3D}^2})(1-\frac{%
\omega_i^2}{\omega^2})}\,,
\end{align}
with $\mathbf{q}=\mathbf{k}-\mathbf{k'}$.  In
Eq.~(\ref{eq:SatOct31152205GMT2009}), $q_{\mathrm{TF}}$ is the 
Thomas-Fermi screening parameter and~$\omega _{\mathrm{ph}}$ is the
phonon dispersion.  The first term between the brackets is bare Coulomb
repulsion, and the second term, which is frequency dependent, follows from
the perturbative coupling to the lattice, tracing out the phonons.

The physical meaning of Eq.~(\ref{eq:SatOct31152205GMT2009}) is at the
heart of the phonon-mediated mechanism. The interaction is $\omega $
dependent, which means, in Fourier transform, time dependent. This
reflects the famous retardation effect in superconductivity. This
effect is based on the strong difference between the electron Fermi
velocity in metals and the sound velocity. Roughly speaking, a fast
electron from the Fermi surface creates a slow phonon and goes
away. After some time, another fast electron arrives and absorbs the
slow phonon. The average distance between these two electrons remains
of the order of 100nm, the distance at which the screened Coulomb
repulsion can be safely neglected. Due to the retardation effect, a
weak phonon-mediated attraction of electrons wins over their Coulomb
repulsion and provides formation of Cooper pairs at low enough
temperatures. %

\section{Interaction hamiltonian}
\label{sec:TueMay25181751BST2010}

A sketch of the structure we propose appears in
Fig.~\ref{fig:ThuApr8130645BST2010}(c). A QW is doped negatively (with
density of electrons $n$) and is put in contact with another QW where
excitons are formed and are stable. The first (lower) QW hosts the
2DEG which is to undergo superconductivity while the second (upper) QW
hosts the condensate of excitons that is to mediate it. This structure
can also be placed at the maximum of the optical field confined in a
microcavity~\cite{laussy10a} (formed by two Bragg mirrors facing each
other). The latter configuration also has the advantage that two
excitonic-QWs can sandwich the 2DEG-QW since in this case the
condensate is delocalised in the entire structure thanks to the photon
fraction of the polariton. This allows an increase by a factor~2, or
more if the multi-layer structure is further repeated, of the
densities achievable in this system. The drawback of microcavities is
a short radiative lifetime of polaritons. In order to maintain the
polariton condensate one needs to pump it resonantly by a high
intensity laser, which may lead to undesirable heating of the
system. The alternative system which we consider is a condensate of
long-living spatially indirect excitons in coupled QWs, separated by a
thin but high barrier from a QW containing a two-dimensional electron
gas (2DEG). This system is simpler to realise as it does not contain
the optical cavity. The condensate of indirect excitons may be
maintained by a low intensity optical excitation. Moreover, indirect
excitons in coupled QWs possess very significant dipole moments which
strengthens their interaction with electrons from a 2DEG. However, 
Bose condensates of indirect excitons have been found only at
temperatures below 1K~\cite{butov02a}, while condensates of
exciton-polaritons are now routinely produced at room
temperature~\cite{butte10a}. In the following, we compare the
advantages of the two respective schemes.  Most of the underlying model
applies to both equally. We focus more particularly on the microcavity
system since it is more general. The exciton system is recovered by
decoupling the photons.

In Fig.~\ref{fig:ThuApr8130645BST2010}(a), the exciton
(resp.~polariton) dispersion is shown in dashed purple (resp.~solid
blue). The difference between the bare polariton dispersion (thick
blue) and the bogolon dispersion (thin blue) is very small over the
range of exchanged momenta of interest (of the order of the Fermi
wavevector). The condensate in both cases is at~$k=0$.  Scattered
particles at any wavevector between $-2k_{\mathrm{F}}$ and $2k_{%
  \mathrm{F}}$ mediate electron-electron interactions on the Fermi
sea, as sketched on Fig.~\ref{fig:ThuApr8130645BST2010}(b). 
\begin{figure}[tbp]
\centering
\includegraphics[width=\linewidth]{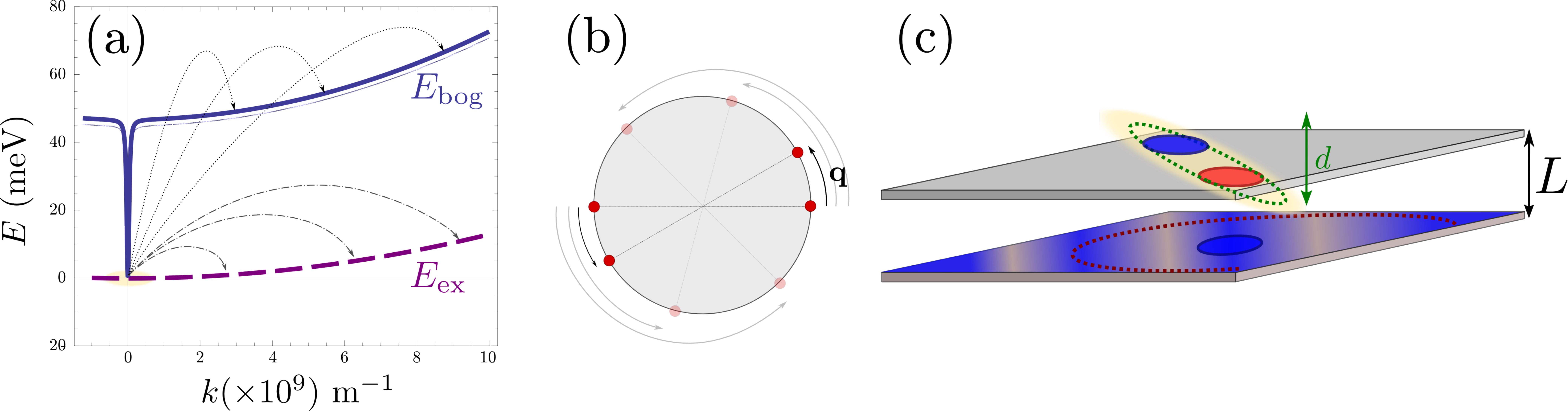}
\caption{One of the possible designs to evidence exciton-polariton
  mediated superconductivity. (a) Polariton (thick blue solid line)
  and exciton (purple dashed line) dispersions with schematic
  representation of scattering of the boson that mediates the
  interaction between electrons. Also shown in thin blue solid line is
  the renormalised dispersion~$E_\mathrm{bog}$, essentially identical
  to $E_\mathrm{pol}$. (b) Corresponding scattering of two electrons
  on the surface of the Fermi sea, exchanging momentum~$\mathbf{q}$
  through scattering of a boson in panel~(a). (c) detail of the
  sandwich structure, showing the doped well (below) containing free
  electrons and the well hosting the condensate (above) of excitons or
  of exciton-polaritons. This layer can be sandwiched in different
  ways.}
\label{fig:ThuApr8130645BST2010}
\end{figure}
The model microscopic Hamiltonian is taken as:~\cite{laussy10a} 
\end{subequations}
\begin{multline}  \label{eq:ThuMar25162758GMT2010}
H=\sum_{\mathbf{k}}\left[E_{\mathrm{pol}}(\mathbf{k}){a_{\mathbf{k}%
}^{\dagger }}a_{\mathbf{k}}+E_{\mathrm{el}}(\mathbf{k}){\sigma_{\mathbf{k}%
}^{\dagger }}\sigma _{\mathbf{k}}\right]+ \\
+\sum_{\mathbf{k}_{1},\mathbf{k}_{2},\mathbf{q}}\Bigg[ V_\mathrm{C}(\mathbf{q%
}){\sigma_{\mathbf{k}_1+\mathbf{q}}^{\dagger}}{\sigma_{\mathbf{k}_2-\mathbf{q%
}}^{\dagger}}\sigma_{\mathbf{k}_1}\sigma_{\mathbf{k}_2,} +XV_\mathrm{X}(%
\mathbf{q}){\sigma _{\mathbf{k}_{1}}^{\dagger }}\sigma _{\mathbf{k}_{1}+%
\mathbf{q}}{a_{\mathbf{k}_{2}+\mathbf{q}}^{\dagger }}a_{\mathbf{k}_{2}} +U{%
a_{\mathbf{k}_{1}}^{\dagger }}{a_{\mathbf{k}_{2}+\mathbf{q}}^{\dagger}}a_{%
\mathbf{k}_{1}+\mathbf{q}}a_{\mathbf{k}_{2}}\Bigg]\,,
\end{multline}
with $E_\mathrm{pol}(\mathbf{k})$ and $E_\mathrm{el}(\mathbf{k})$ the
polariton and 2DEG dispersions for the in-plane wavevector~$\mathbf{k}$,
respectively. In the exciton case (without the microcavity), it suffices to
replace $E_\mathrm{pol}$ with $E_\mathrm{ex}$ in the above. $V_\mathrm{X}$
is the electron-polariton interaction, $U$ the polariton-polariton
interaction and $V_\mathrm{C}$ the electron-electron repulsion. We now
consider these terms in turn. Parameters assumed are listed in
table~\ref{tab:MonNov29002411CET2010}.

\begin{table}[hbtp]
  \centering
  \begin{tabular}{c|l|c}
    Parameter & Meaning & Value \\
    \hline
    \hline
    $\epsilon$ & Permittivity & $7\epsilon_0\approx\unit{6.2}\ampere\second/(\meter\volt)$ \\
    $\beta_e$ & Electron reduced mass & $\frac{0.22}{0.22+1.25}\approx1.15$\\
    $\beta_h$ & Hole reduced mass & $\frac{1.25}{0.22+1.25}\approx0.85$\\
    $L$ & Distance between wells & \unit{5}\nano\meter \\
    $\kappa$ & Coulomb screening length & $\approx$\unit{1.2\times10^9}\meter$^{-1}$ \\
    $m_\mathrm{x}$ & Exciton mass & $(0.22+1.25)m_e\approx\unit{1.3\times10^{-30}}\kilo\gram$ \\
    $m_\mathrm{c}$ & Photon mass & $10^{-5}m_e\approx\unit{9.1\times10^{-36}}\kilo\gram$\\
    $2g$ & Rabi splitting & $\unit{45}\milli\electronvolt$ [0]\\
    $X$ & Hopfield coefficient (exciton weight) & $1/\sqrt{2}$ [1]\\
    $k_\mathrm{F}$ & Fermi wavevector & $\unit{5\times10^8}\meter^{-1}$\\
    $a_\mathrm{B}$ & Exciton Bohr radius & $\unit{1.98\times10^{-9}}\meter$ \\
    $R_y$ & Exciton Rydberg & $\unit{32}\milli\electronvolt$ \\
    $l$ & Dipole moment & $\unit{4}\nano\meter$\\
  \end{tabular}
  \caption{Parameters used in the numerical simulations. In square brackets, the value for the exciton case when the parameter differs from the polariton case, otherwise parameters have been taken the same for comparison.}
  \label{tab:MonNov29002411CET2010}
\end{table}

\subsection{Electron-electron interaction}

\begin{figure}[hbpt]
\centering
\includegraphics[width=.7\linewidth]{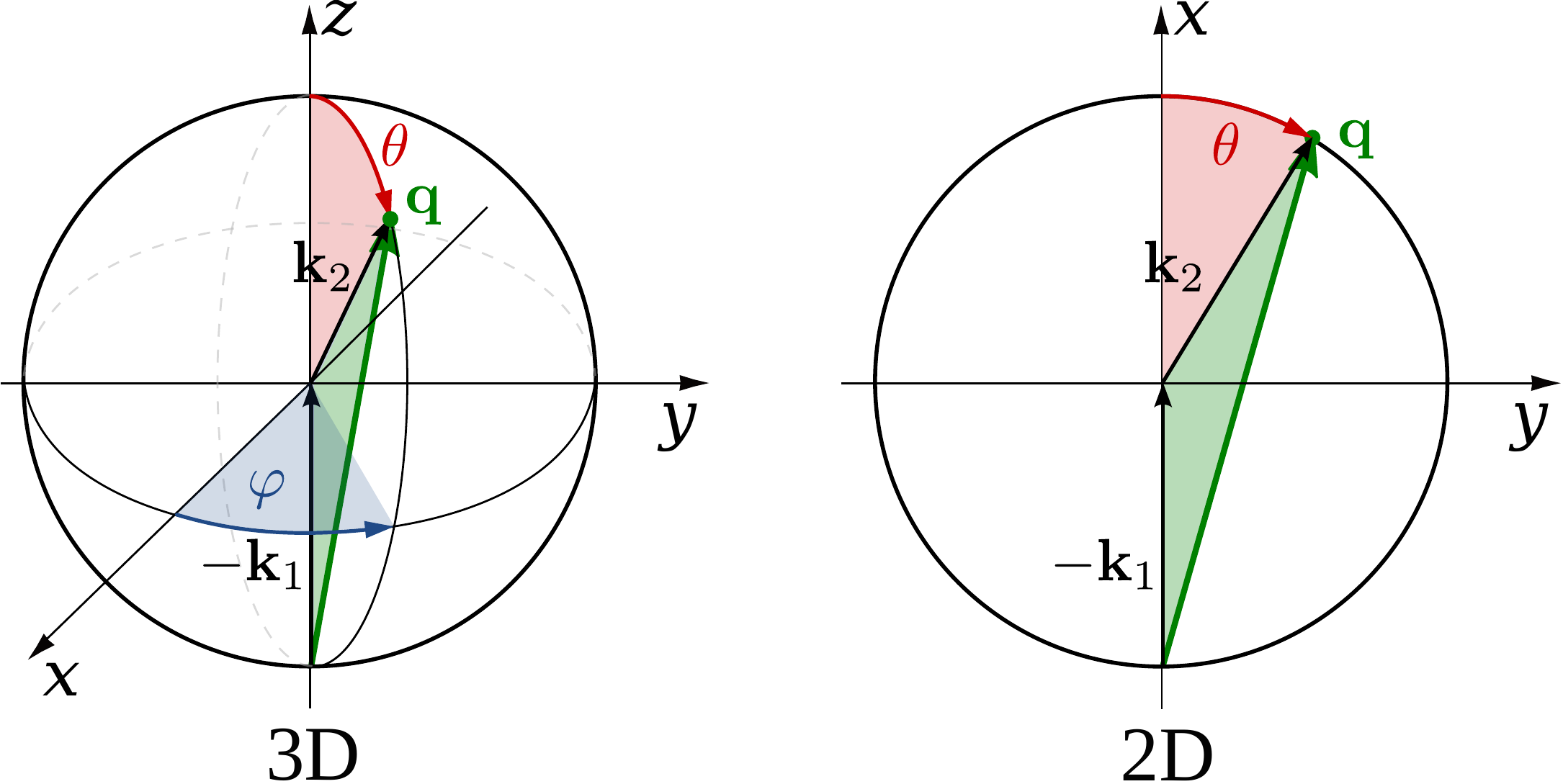}
\caption{Average over the Fermi Surface, in 3D and 2D.}
\label{fig:ThuOct29173853GMT2009}
\end{figure}

In the original BCS mechanism, electron-electron repulsion is either
neglected altogether or overcome by the attractive mechanism and not
manifested outside of the attractive window. We take it into account here
since it is a detrimental factor for binding and most of our concern for
experimental realization is to optimize this value. The full form of the
potential is given by the Yukawa potential: 
\begin{equation}
V_{\mathrm{C}}(\mathbf{q})=\frac{e^{2}}{2\epsilon A}\frac{1}{|\mathbf{q}%
|+\kappa }  \label{eq:FriMar26140915GMT2010}
\end{equation}%
with screening constant~$\kappa $. We get rid of the momentum dependence by
averaging the potential $V_{\mathrm{eff}}(\omega ,\mathbf{q})$ over the
Fermi surface (FS), where: 
\begin{equation}
\mathbf{q}=\mathbf{k_{1}}-\mathbf{k_{2}}\,  \label{eq:ThuOct29173054GMT2009}
\end{equation}%
and~$\mathbf{k}_{1,2}$ are the momenta of the two interacting electrons on
the FS (so that, in particular, $k_{1}=k_{2}=k_{\mathrm{F}}$), as shown in
Fig.~\ref{fig:ThuOct29173853GMT2009}. In 3D, the FS is the surface of a
sphere, while in 2D, it is a circle (we shall speak of surface in both
cases). The vector difference is therefore joining the two end-points on the
surface. If the potential has spherical symmetry ($V(\mathbf{q})=V(q)$), the
average of all two vectors on the FS reduces to that where~$\mathbf{q}$ is
pinned at one point of the surface (the south pole in Fig.~\ref%
{fig:ThuOct29173853GMT2009}) and runs overs the FS. This average, in the
particular choice of Fig.~\ref{fig:ThuOct29173853GMT2009}, is the usual
polar integration with~$\mathbf{k}_{2}$ describing the surface as $\theta $
(and~$\phi $ in 3D) are varied, with: 
\begin{equation}
q^{2}=2k_{\mathrm{F}}^{2}(1+\cos \theta )\,,
\label{eq:ThuOct29175945GMT2009}
\end{equation}%
from Al Kashi's theorem, so that the average potential $\bar{V}_{\mathrm{eff}%
}$ reads, in 3D: 
\begin{equation}
\bar{V}_{\mathrm{eff}}(\omega )=\int_{0}^{2\pi }\int_{0}^{\pi }V_{\mathrm{eff%
}}(q,\omega )k_{\mathrm{F}}^{2}\sin \theta \,d\theta d\varphi /\mathcal{N}\,,
\label{eq:ThuOct29180136GMT2009}
\end{equation}%
where~$\mathcal{N}$ is the normalization, i.e., the same integral where~$V_{%
\mathrm{eff}}$ is replaced by unity. This gives, in 3D: 
\begin{align}
\bar{V}_{\mathrm{eff}}(\omega )& =\frac{1}{2}\int_{-1}^{1}V_{\mathrm{eff}}(%
\sqrt{2k_{\mathrm{F}}^{2}(1+\cos \theta )},\omega )d\cos \theta \,,
\label{eq:ThuOct29182211GMT2009} \\
& =\frac{1}{4}\int_{0}^{4}V_{\mathrm{eff}}(k_{\mathrm{F}}\sqrt{\vartheta }%
,\omega )d\vartheta \,,  \label{eq:SunNov1133021GMT2009}
\end{align}%
where we integrate over $\vartheta =2(1+\cos \theta )$ since this is a
natural variable in Eq.~(\ref{eq:ThuOct29175945GMT2009}), and, in 2D: 
\begin{equation}
\bar{V}_{\mathrm{eff}}(\omega )=\frac{1}{2\pi }\int_{0}^{2\pi }V_{\mathrm{eff%
}}(\sqrt{2k_{\mathrm{F}}^{2}(1+\cos \theta )},\omega )d\theta \,.
\label{eq:FriOct30115236GMT2009}
\end{equation}

\begin{figure}[tbp]
\centering
\includegraphics[width=.5\linewidth]{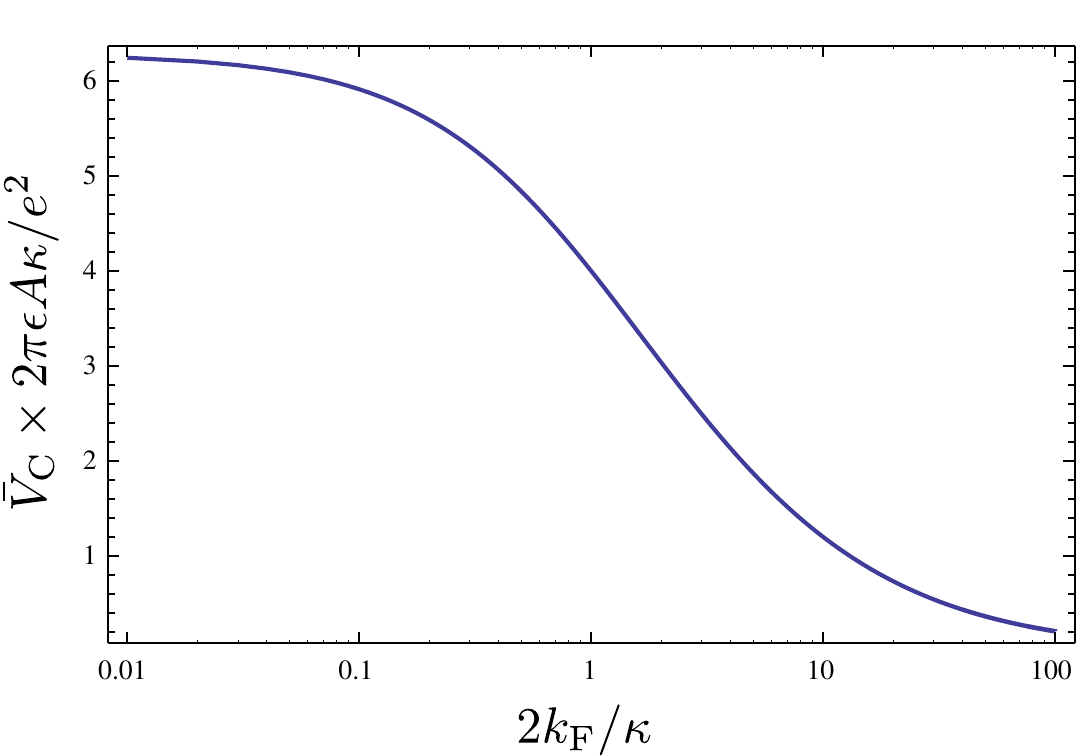}
\caption{Average electron-electron repulsion (in natural units) in the QW.
This should be made as small as possible, which is obtained for higher Fermi
wavevectors (for a given screening length).}
\label{fig:SatMar27224820GMT2010}
\end{figure}

Our system is 2D, in which case, from
Eqs.~(\ref{eq:FriMar26140915GMT2010})
and~(\ref{eq:FriOct30115236GMT2009}):
\begin{align}  \label{eq:MonDec7173053GMT2009}
\bar V_\mathrm{C}=\frac{e^2}{4\pi\epsilon A}\int_0^{2\pi}\frac{d\theta}{%
\sqrt{2k_\mathrm{F}^2(1+\cos\theta)}+\kappa}&=\frac{e^2}{2\pi\epsilon A}%
\frac{\pi-2\arctan\frac{2k_\mathrm{F}}{\sqrt{\kappa^2-4k_\mathrm{F}^2}}}{%
\sqrt{\kappa^2-4k_\mathrm{F}^2}}\quad & \text{if $\kappa\ge2k_\mathrm{F}$},
\\
&=\frac{e^2}{2\pi\epsilon A} \frac{\ln\left|\frac{1+2k_\mathrm{F}\Big/\sqrt{%
4k_\mathrm{F}^2-\kappa^2}}{1-2k_\mathrm{F}\Big/\sqrt{4k_\mathrm{F}^2-\kappa^2%
}}\right|}{\sqrt{4k_\mathrm{F}^2-\kappa^2}}\quad & \text{if $\kappa\le2k_%
\mathrm{F}$}.
\end{align}

This is plotted in Fig.~\ref{fig:SatMar27224820GMT2010}. Since we are trying
to maximize attraction, that is, minimize repulsion, systems with small
screening length and large wavevectors should be favored (but these
parameters play critically on other aspects of the mechanism and the optimum
is not compulsorily $k_\mathrm{F}/\kappa\gg1$).

\subsection{Electron-exciton interaction}

The electron-exciton or exciton-polariton interaction is one of the
most important ingredients of the mechanism, as it ultimately
determines the shape of the effective potential. In the microcavity,
an electron (from the 2DEG) interacts with a polariton (from the
condensate) through its excitonic component, so this is really the
electron-exciton interaction that is to be computed, weighted by the Hopfield
coefficient (the excitonic fraction) $X$. Let us consider, therefore,
the scattering of an electron in one of the parallel QW, separated by
a distance $L$ from the QW with excitons. The matrix element of the
direct interaction between excitons and electrons reads:
\begin{equation}
V_{X}(\mathbf{q})=\int \Psi _{X}^{\ast }(\mathbf{Q},\mathbf{r}_{e},\mathbf{r}%
_{h})\Psi ^{\ast }(\mathbf{k},\mathbf{r}_{1})V(\mathbf{r}_{1},\mathbf{r}_{e},%
\mathbf{r}_{h})\Psi _{X}^{\ast }(\mathbf{Q},\mathbf{r}_{e},\mathbf{r}%
_{h})\Psi ^{\ast }(\mathbf{k},\mathbf{r}_{1})d\mathbf{r}_{1}d\mathbf{r}_{e}d%
\mathbf{r}_{h}
\end{equation}%
where $\mathbf{r}_{1},\mathbf{r}_{e},\mathbf{r}_{h}$ correspond to the 2D
coordinates of the 2DEG electron, the exciton electron and the exciton hole
respectively. The 2DEG electron is described by a plane wave while the
electron/hole in the condensate are assumed to be in the 1s bound state with plane
wave center-of-mass motion:
\begin{subequations}
\begin{align}
\Psi (\mathbf{q},\mathbf{r}_{1})& =\frac{1}{\sqrt{A}}e^{i\mathbf{kr}_{1}}, \\
\Psi _{X}^{\ast }(\mathbf{Q},\mathbf{r}_{e},\mathbf{r}_{h})& =\sqrt{\frac{2}{%
\pi A}}\frac{1}{a_{B}}U_{e}\left( z_{e}\right) U_{h}\left( z_{h}\right) e^{i%
\mathbf{Q}\cdot (\beta _{e}\mathbf{r}_{e}+\beta _{h}\mathbf{r}_{h})}e^{-|%
\mathbf{r}_{e}-\mathbf{r}_{h}|/a_{B}}=\sqrt{\frac{2}{\pi A}}\frac{%
U_{e}\left( z_{e}\right) U_{h}\left( z_{h}\right) }{a_{B}}e^{i\mathbf{Q}%
\cdot \mathbf{R}_{X}}e^{-r_{X}/a_{B}}\,,
\end{align}%
where $\mathbf{R}_{X}=\beta _{e}\mathbf{r}_{e}+\beta
_{h}\mathbf{r}_{h},%
\mathbf{r}_{X}=\mathbf{r}_{e}-\mathbf{r}_{h}$ are in-plane coordinates
of the center of mass of the exciton and relative coordinate of
electron and hole in the exciton, $U_{e}\left( z_{e}\right) $ and
$U_{h}\left( z_{h}\right) $ are normal to the QW plane electron and
hole envelope functions, respectively.  We also consider the existence
of a dipole moment $d$ for the exciton, which can be intrinsic to the
structure, because of spatial separation of electrons and holes in
coupled QWs, or (in the case of microcavities) be induced by an internal
piezo-electric field, or result from an externally applied electric
field. To account for all these possibilities, one can consider the
layers of electrons and holes in the exciton shifted in the
$z$-direction with respect to the position of the center of mass by
a distance $l=d/e\ll L$. The matrix element of the interaction is
then computed to be:

\end{subequations}
\begin{subequations}
\begin{align}
V_{dir}(q)=&\frac{e^2}{2\epsilon A}\frac{e^{-qL}}{q}\left\{\frac{1}{\left[%
1+(\beta_e q a_B/2)^2\right]^{3/2}}-\frac{1}{\left[1+(\beta_h q a_B/2)^2%
\right]^{3/2}}\right\}  \label{eq:ThuMay27134002BST2010} \\
+&\frac{ed}{2\epsilon A}e^{-qL}\left\{\frac{\beta_e}{\left[1+(\beta_e q
a_B/2)^2\right]^{3/2}}+\frac{\beta_h}{\left[1+(\beta_h q a_B/2)^2\right]%
^{3/2}}\right\}  \label{eq:ThuMay27134007BST2010}
\end{align}
where~Eq.~(\ref{eq:ThuMay27134002BST2010}) is the direct
electron-exciton interaction that exists even in the absence of a
dipole moment of the exciton, and Eq.~(\ref{eq:ThuMay27134007BST2010})
is the dipolar interaction.  The direct interaction vanishes at small
exchanged momenta, while the dipolar-induced one assumes its maximum
value here of ${2d}%
/(2\epsilon_0\epsilon A)$. Overall, the dipolar interaction is
naturally much larger than the direct one, since the exciton is
electrically neutral.%
\footnote{%
  Note at this point that there is an error in Ref~\cite{laussy10a}
  where only the direct exciton interaction has been taken into
  account with an incorrect power in the parenthesis, which led to an
  expression similar to the dipolar interaction. The direct
  interaction by itself turns out to be too small to evidence
  superconductivity with the parameters chosen in Ref.~%
  \cite{laussy10a}, therefore a dipole moment should be induced in
  this case, say by applying an external electric field, to restore
  the effect.} These facts are summarized in
Fig.~\ref{fig:ThuMay27143618BST2010}.

\begin{figure}[tbp]
\centering
\includegraphics[width=.6\linewidth]{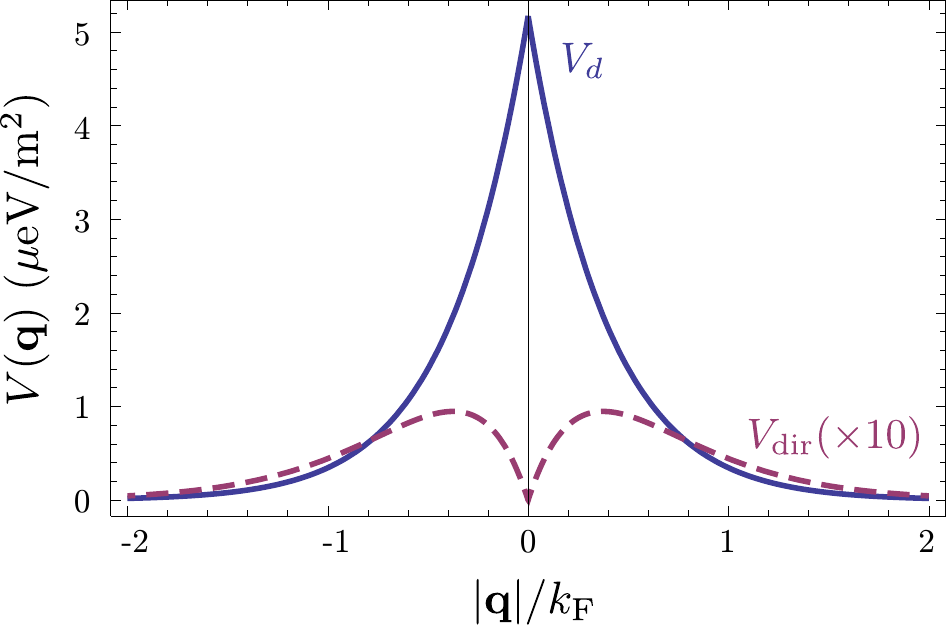}
\caption{Electron-exciton interaction in the geometry of Fig.~\protect\ref%
{fig:ThuApr8130645BST2010}, decomposed as the direct interaction (dashed purple)
and dipolar interaction (solid blue) when the exciton is induced with a
dipole moment~$d$. The latter is both much larger and maximum at zero
exchanged momentum.}
\label{fig:ThuMay27143618BST2010}
\end{figure}

\subsection{Polariton-polariton interaction}

We treat the polariton-polariton interaction within the $s$-wave
scattering approximation with strength $%
U=6a_{\mathrm{B}}^{2}R_{y}X^{2}/A$ (where $a_{\mathrm{B}}$ is the
exciton Bohr radius, $R_{y}$ the exciton binding energy and $A$ the
normalization area. $X$ is the exciton Hopfield coefficient, the square  
of which quantifies the exciton fraction in the exciton-polariton
condensate)~\cite{tassone99a}. For interaction between bare excitons,
$X=1$. Exciton-exciton (polariton-polariton) interactions are
repulsive, in general. They result in linearization of the elementary
excitation spectra, the Bogoliubov dispersion, but its role is not
crucial to the mechanism of superconductivity we discuss, since at the
wavevectors of interest, the changes brought by this term are very
small compared to the kinetic energy of non-interacting excitons
(exciton-polaritons).

\subsection{Effective interaction}

We now proceed to bring our microscopic model towards a form suited to
study Cooper-pairing and superconductivity, that is, we apply the
canonical Fr\"ohlich transformation that will result in an effective
BCS hamiltonian.  Just as in the case of phonons, we start by getting
rid of polaritons. We assume a condensate is formed with mean
population $N_{0}$. We do not consider which mechanism, coherent or
incoherent, is responsible for creating and maintaining this state. We
do assume, however, it is coherent and with a definite phase, so that
we can apply the mean-field approximation $%
a_{\mathbf{k}_{1}+\mathbf{q}}^{\dagger }a_{\mathbf{k}_{1}}\approx
\langle a_{%
  \mathbf{k}_{1}+\mathbf{q}}^{\dagger }\rangle
a_{\mathbf{k}_{1}}+a_{\mathbf{k}%
  _{1}+\mathbf{q}}^{\dagger }\langle a_{\mathbf{k}_{1}}\rangle $
and~$\langle a_{\mathbf{k}}\rangle \approx \sqrt{N_{0}A}\delta
_{\mathbf{k},\mathbf{0}}$ with $N_{0}$ the density of polaritons in
the condensate. This allows us to obtain the following expression for
the Hamiltonian, after diagonalizing the polariton part by means of a
Bogoliubov transformation (that leaves the free propagation of
electrons and their direct interaction, $H_{\mathrm{C}}$, invariant):
\end{subequations}
\begin{equation}
H=\sum_{\mathbf{k}}E_{\mathrm{el}}(\mathbf{k}){\sigma _{\mathbf{k}}^{\dagger
}}\sigma _{\mathbf{k}}+\sum_{\mathbf{k}}E_{\mathrm{bog}}(\mathbf{k}){b_{%
\mathbf{k}}^{\dagger }}b_{\mathbf{k}}+H_{\mathrm{C}}+\sum_{\mathbf{k},%
\mathbf{q}}M(\mathbf{q}){\sigma _{\mathbf{k}}^{\dagger }}\sigma _{\mathbf{k}+%
\mathbf{q}}({b_{-\mathbf{q}}^{\dagger }}+b_{\mathbf{q}})\,
\label{Hamiltonian_3}
\end{equation}%
where $E_{\mathrm{bog}}(\mathbf{k})$ describes the dispersion of the
elementary excitations (bogolons) of the interacting Bose gas, which is very
close to a parabolic exciton dispersion at large~$k$: 
\begin{equation}
E_{\mathrm{bog}}(\mathbf{k})=\sqrt{\tilde{E}_{\mathrm{pol}}(\mathbf{k})(%
\tilde{E}_{\mathrm{pol}}(\mathbf{k})+2UN_{0}A)}\,,
\end{equation}%
where $\tilde{E}_{\mathrm{pol}}(\mathbf{k})\equiv E_{\mathrm{pol}}(\mathbf{k}%
)-E_{\mathrm{pol}}(\mathbf{0})$ and with the renormalized bogolon-electron
interaction strength: 
\begin{equation}
M(\mathbf{q})=\sqrt{N_{0}A}XV_{\mathrm{X}}(\mathbf{q})\sqrt{\frac{E_{\mathrm{%
bog}}(\mathbf{\ q})-\tilde{E}_{\mathrm{pol}}(\mathbf{q})}{2UN_{0}A-E_{%
\mathrm{bog}}(\mathbf{q})+\tilde{E}_{\mathrm{pol}}(\mathbf{q})}}\,.
\label{Frohlich}
\end{equation}

The last term of Eq.~(\ref{Hamiltonian_3}) coincides with the
Fr\"ohlich electron-phonon interaction Hamiltonian, which allows us to
write an effective Hamiltonian for the bogolon-mediated
electron-electron interaction. This results in an effective
interaction between electrons, of the type $\sum_{\mathbf{k}_{1},
  \mathbf{k}_{2},\mathbf{q}}V_{\mathrm{eff}}(%
\mathbf{q},\omega ){\sigma _{ \mathbf{k}_{1}}^{\dagger }}\sigma
_{\mathbf{k}%
  _{1}+\mathbf{q}}{\sigma}_{\mathbf{k}_{2}+\mathbf{q}}^{\dagger}\sigma
_{%
  \mathbf{k}_{2}}$. The effective interaction strength reads
$V_{\mathrm{eff}}(%
\mathbf{q},\omega
)=V_\mathrm{C}(\mathbf{q})+V_\mathrm{A}(\mathbf{q},\omega)$%
, with:
\begin{equation}
V_\mathrm{A}(\mathbf{q},\omega)=\frac{2M(\mathbf{q})^{2}E_{\mathrm{bog}}(%
\mathbf{q})}{(\hbar \omega )^{2}-E_{\mathrm{bog}}(\mathbf{q})^{2}}\,.
\label{V_eff}
\end{equation}

\begin{figure}[tbp]
\centering
\includegraphics[width=\linewidth]{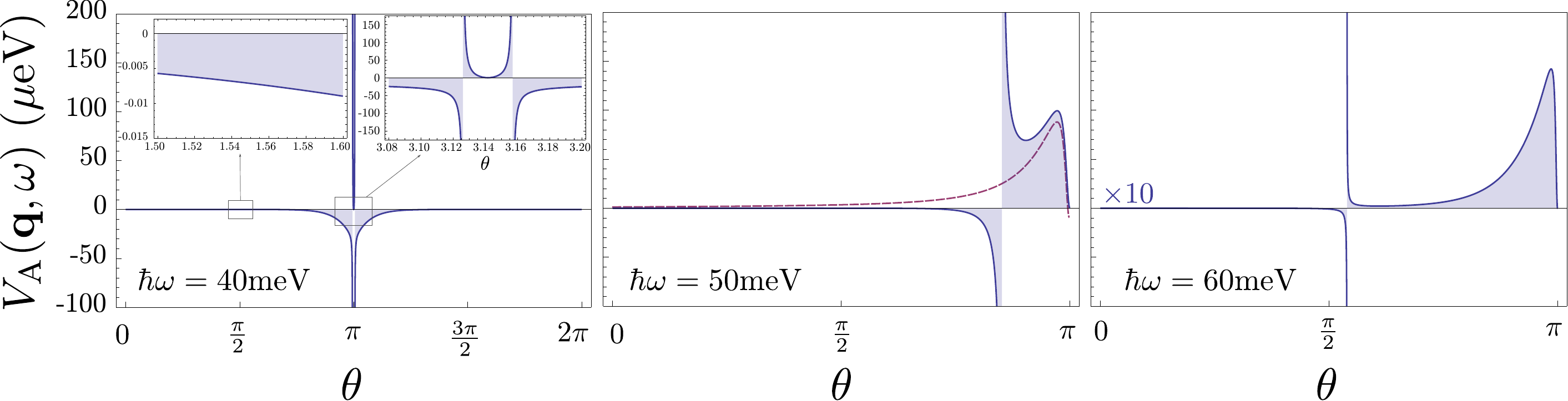}
\caption{Effective electron-electron interaction as a function of the
exchanged momentum $q=\protect\sqrt{2}k_\mathrm{F}\protect\sqrt{1+\cos%
\protect\theta}$ on the Fermi sea at energies $\hbar\protect\omega=40$, 50
and 60meV, respectively. The potential is symmetric around~$\protect\pi$. In
the first case, the whole dependency (over~$[0,2\protect\pi]$) is shown,
with a first zoom in the inset showing two poles and another the attractive
region that is of small amplitude but extends over a large range. With
increasing energies, the poles recedes towards smaller values of $\protect%
\theta$ with a dominating effect of the repulsive (positive) energy. In the
central panel, in dashed purple, the regularized potential is superimposed.}
\label{fig:ThuMay27163615BST2010}
\end{figure}

Equation~(\ref{V_eff}) recovers the boson-mediated interaction
potential obtained for a Bose-Fermi mixture of cold atomic
gases~\cite{bijlsma00a}, in the limit of vanishing exchanged
wavevectors. It describes the BEC\ induced attraction between
electrons. Remarkably, it increases linearly with the condensate
density $N_{0}$. This represents an important advantage of this
mechanism of superconductivity with respect to the earlier proposals
of exciton-mediated superconductivity~\cite%
{little64a,ginzburg_book09a,allender73a}, as the strength of Cooper
coupling can be directly controlled by optical pumping of the
exciton-polariton condensate. The attractive potential is displayed
for various exchanged energies in
Fig.~\ref{fig:ThuMay27163615BST2010}, as a function of the exchanged
momentum expressed directly through the angle~$\theta$ defined in the
Fermi circle (cf.~Fig.~\ref{fig:ThuOct29173853GMT2009}). As commented
earlier, the negative part corresponds to attraction, and the
potential alternates between repulsive and attractive character,
obtained at different exchanged momenta. We do not want to keep track
of such complicated wavevector dependence, and therefore will average
the interaction over the Fermi sea. A notable feature is, for most
values of $\omega$, the presence of a pole $\theta_0$, where
$E_\mathrm{bog}(\sqrt{2k_\mathrm{F}%
  ^2(1+\cos\theta_0}))=\hbar\omega$. As is seen in the figure,
$\theta_0$ separates the attractive part from the repulsive part. The
average will bring the additional convenience of cancelling such
divergencies. We note as well that the spectrum of excitations of the
exciton BEC may be changed in the presence of the electron gas, so
that their eventual dispersion may be different
$E_\mathrm{bog}$~\cite{shelykh10a}. This has no effect on the Cooper
pairing of electrons which we discuss here.

\begin{figure}[tbp]
\centering
\includegraphics[width=\linewidth]{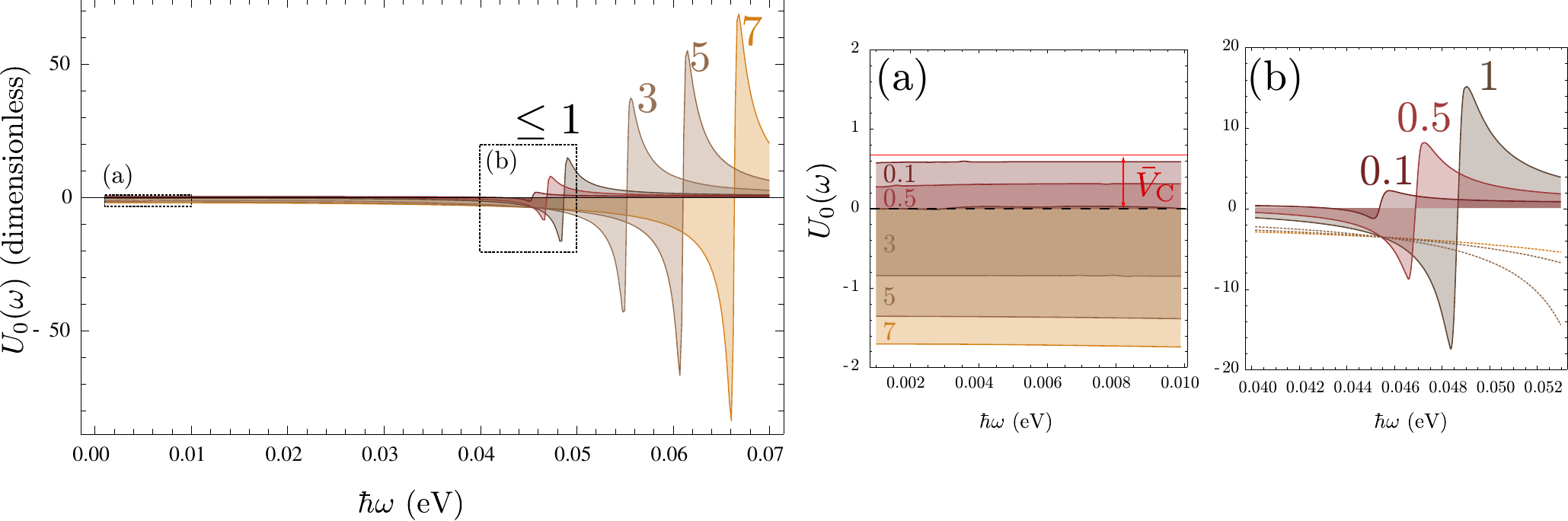}
\caption{(Colour online) Exchanged-momentum averaged interaction $U_0(%
\protect\omega)$ between electrons of the 2DEG, as $N_0$ is increased. It is
dimensionless and is attractive (resp.~repulsive) when negative
(resp.~positive). Densities $N_0$ are shown in units of $10^{12}/$cm$^{2}$.
In insets (a) and~(b), a zoom of the regions delineated on the central
(left) panel, firstly in the small energies (long time) range, where the
attraction is seen to be repulsive for smaller densities, because of the
Coulomb repulsion (shown in red), and secondly for the densities $\approx10^{12}/$%
cm$^{2}$ in the area where the character of the interaction changes abruptly
from attractive to repulsive (higher densities are shown in dotted lines).}
\label{fig:SatMay29135948BST2010}
\end{figure}

We therefore wish to perform the average 
\begin{equation}  \label{eq:WedJan27191839GMT2010}
\bar V_\mathrm{A}=\int_0^{2\pi}V_\mathrm{A}(q,\omega)\,d\theta\,,
\end{equation}
where $q=\sqrt{2k_\mathrm{F}^2(1+\cos\theta)}$, as seen
previously. Since $V_\mathrm{%
  A}$ is symmetric around~$\pi$ we perform the integral $\int_0^\pi$
only. The integral would be easily computed numerically if there were
no pole.  There are dedicated numerical methods to compute
principal values numerically~\cite{thompson98a,noble00a}, but in our
case, since the pole is first order, it is enough to isolate it
analytically by defining
\begin{equation}  \label{eq:WedJan27192330GMT2010}
f(\theta)=(\theta-\theta_0)V_\mathrm{A}(\mathbf{q}(\theta),\omega)\,,
\end{equation}
so that then 
\begin{equation}  \label{eq:WedJan27192632GMT2010}
\bar V_\mathrm{A}=\int_0^{2\pi}\frac{f(\theta)-f(\theta_0)}{\theta-\theta_0}%
\,d\theta+f(\theta_0)\ln\frac{2\pi-\theta_0}{\theta_0}
\end{equation}
where $f(\theta_0)=\lim_{\theta\rightarrow\theta_0}f(\theta)$ and the first
integral is regular (it is shown in Fig.~\ref{fig:ThuMay27163615BST2010} in
dashed purple). The integration (average) is then straightforward and
produces the results shown in Fig.~(\ref{fig:SatMay29135948BST2010}).

\begin{figure}[tbp]
\centering
\includegraphics[width=\linewidth]{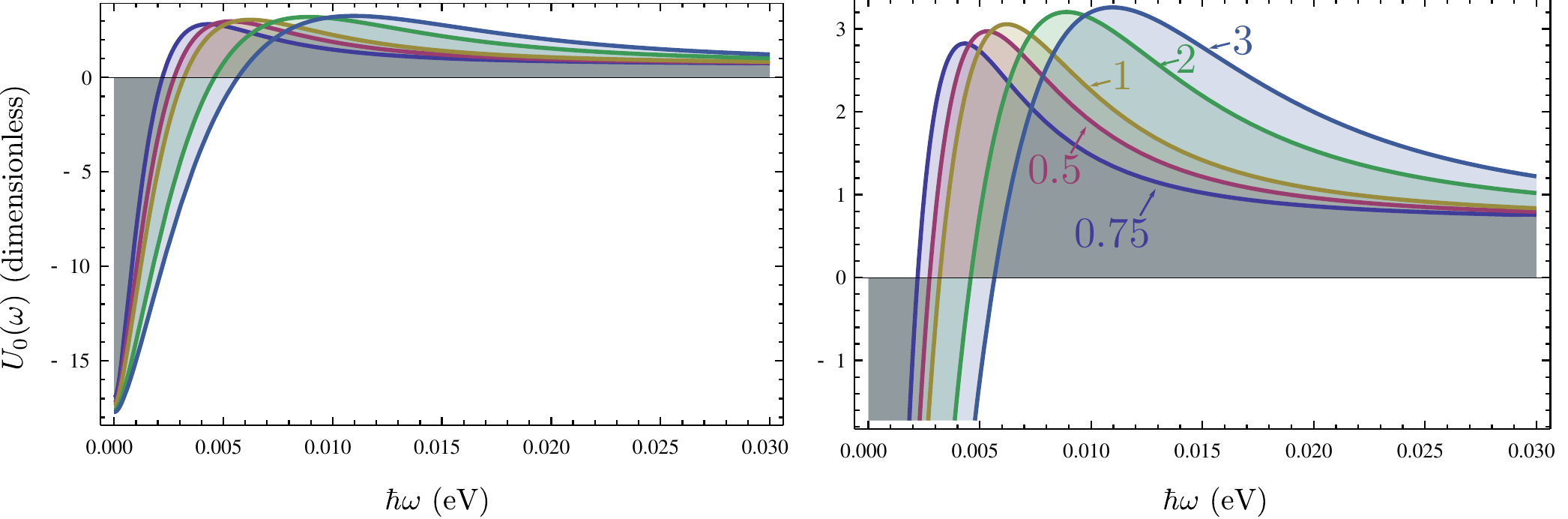}
\caption{Same as Fig.~\protect\ref{fig:SatMay29135948BST2010} (also in units
of $10^{12}/$cm$^{2}$) but for the case of an exciton condensate. A larger
dipole moment $d$ has also been assumed. The potential is different in
character, much closer to Cooper's potential with larger attraction at
longer times.}
\label{fig:SatAug21093037BST2010}
\end{figure}

If we take instead of $E_{\mathrm{pol}}$ a quadratic dispersion for the
excitation 
\begin{equation}
E_{\mathrm{x}}(k)=\frac{\hbar ^{2}k^{2}}{2m_{\mathrm{x}}}
\label{eq:MonFeb1124400GMT2010}
\end{equation}%
and assume all-excitonic interactions, $X=1$, we can consider the same
effect in the absence of a microcavity, relying on a purely excitonic
(rather than polaritonic) BEC.  In this case, the same procedure as
detailed above leads to an effective potential as shown in Fig.~\ref%
{fig:SatAug21093037BST2010}. We kept all parameters the same for
comparison except for the dipole moment, which we have taken three
times as large ($d=12$%
nm). This corresponds to the spatial separation of electrons and holes
in the system of indirect excitons studied by Butov \textit{et al}
\cite{butov02a}. The potential we obtain in the exciton case is very
different in character from the polariton case, and is more closely
related to the Cooper (conventional) shape of a square well, or the
Bogoliubov potential including repulsive elbows.

From these potentials, one can proceed to solve the gap equation.

\section{Gap equation}

\label{sec:TueMay25182122BST2010}

\subsection{Cooper potential}

\begin{figure}[tbp]
\centering
\includegraphics[width=\linewidth]{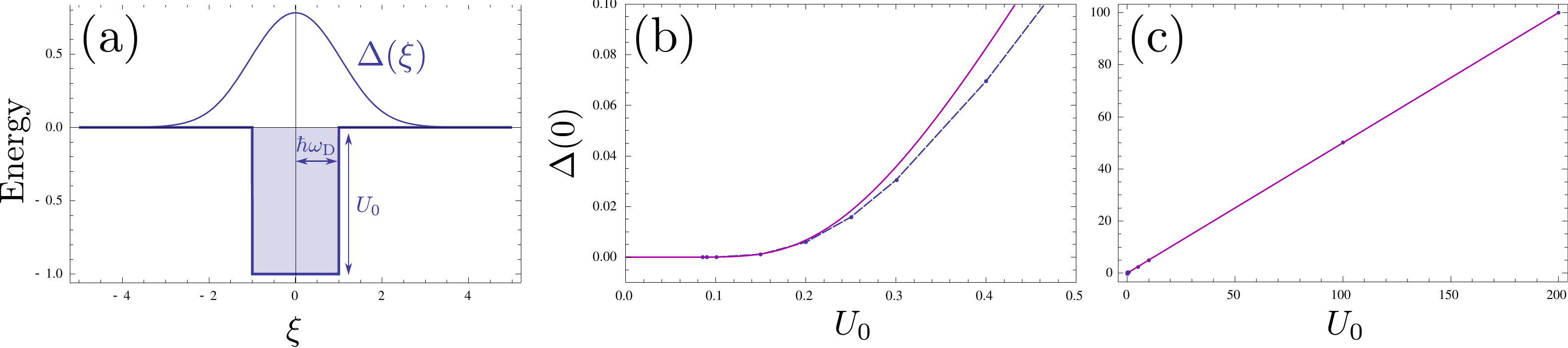}
\caption{Numerical solution to the gap equation with the BCS square well
potential (a). The gap $\Delta(\protect\xi)$ is not stepwise as approximated
in the BCS model, but its value at zero exchanged energy, $\Delta(0)$, is in
close agreement with the analytical expressions, as seen in (b) and (c): $%
\Delta(0)$ as obtained from numerically solving the gap equation (dashed
blue) and from Eq.~(\protect\ref{eq:TueApr6180432BST2010}) (solid magenta),
(b) in the weak-coupling limit when $U_0\ll1$, with small deviations as
coupling is increased, and (c) in the strong-coupling where the agreement
becomes perfect again. Note that $\Delta$ increases linearly with the
coupling strength out of the weak-coupling limit.}
\label{fig:WedApr7114819BST2010}
\end{figure}

The BCS gap equation~(\ref{eq:TueApr6172010BST2010}) is easier to tackle as
a continuous equation: 
\begin{equation}  \label{eq:MonMar29151237BST2010}
\Delta(\xi,T)=-\int_{-\infty}^{\infty}\frac{U_0(\xi-\xi^{\prime
})\Delta(\xi^{\prime },T)\tanh(E/2k_\mathrm{B}T)}{2E}\,d\xi^{\prime }\,,
\end{equation}
where we have also introduced a finite temperature~$T$ from the Fermi-Dirac
distribution of elementary excitations~\cite{degennes_book99a}. With the BCS
approximation of a step potential, the gap equation at zero temperature
simplifies to $\Delta(\xi,T)=-U_0\int_{\xi-\hbar\omega_\mathrm{D}%
}^{\xi+\hbar\omega_\mathrm{D}}{\Delta(\xi^{\prime },T)}/{2E}\,d\xi^{\prime }$%
. If $\Delta\gg\hbar\omega_\mathrm{D}$, the $\xi$ dependence in the integral
boundaries can be neglected (or, coming back to Eq.~(\ref%
{eq:TueApr6172010BST2010}), one sees that in the initial gap equation, $%
\Delta_\mathbf{k}$ is exactly constant if $U_{\mathbf{k}\mathbf{k}^{\prime
}}=U_0$ and is zero otherwise). Thus, we can assume the gap to be of the
form: 
\begin{equation}  \label{eq:TueApr6173939BST2010}
\Delta(\xi)= 
\begin{cases}
\Delta(0) & \text{if $|\xi|\le\hbar\omega_\mathrm{D}$}\,, \\ 
0 & \text{otherwise}\,.%
\end{cases}%
\end{equation}
In this case, simplifying $\Delta(0)$ on both side of Eq.~(\ref%
{eq:MonMar29151237BST2010}), we obtain: 
\begin{equation}  \label{eq:TueApr6180432BST2010}
\Delta(0)=-\hbar\omega_\mathrm{D}/\sinh(-1/U_0)\,.
\end{equation}

Equation (\ref{eq:TueApr6180432BST2010}) is better known as its
approximation when $U_0\ll1$, in which case it takes the form of the famous
BCS gap expression, $\Delta(0)=2\hbar\omega_\mathrm{D}\exp(-1/U_0)$.

Solving exactly the gap equation calls for some numerical method. Equation~(%
\ref{eq:MonMar29151237BST2010}) is a nonlinear integral equation, of the
type studied by Hammerstein, i.e., 
\begin{equation}  \label{eq:MonMar29153626BST2010}
\Delta(\xi,T)=\int K(\xi,\xi^{\prime })f[\xi^{\prime },\Delta(\xi^{\prime
})]\,d\xi^{\prime }
\end{equation}
where in our case $K(x,y)=-U_0(x-y)/2$ and
$f[y,z]=z\tanh(\sqrt{y^2+z^2})%
\big/\sqrt{y^2+z^2}$. There are strong conditions of existence of
nontrivial (nonzero) solutions when
$U_0\le0$~\cite{kitamura63a,vansevenant85a}, however the case when the
kernel $K$ is not positive definite, that is, in presence of
repulsion,\footnote{%
  as well as attraction, since the case of only repulsion admits only
  $\Delta=0 $ as a solution.} has been much less studied
mathematically~\cite%
{zabreiko67a}. Cooper's potential, being always negative, falls in the
category of potentials which admit a unique nontrivial solution, with
a stable numerical technique to obtain it, namely, since the mapping
is contractive, by iterations of the gap equation: an initial
(nonzero) function $\Delta_0$ is used to compute the rhs of Eq.~(\ref%
{eq:MonMar29151237BST2010}), providing $\Delta_1$ which is injected
back until the function converges. The gap computed in this way for a
given Cooper potential is shown in
Fig.~\ref{fig:WedApr7114819BST2010}(a). As one can see, the BCS
approximation~(\ref{eq:TueApr6173939BST2010}) remains a rather coarse
approximation, since the gap turns out in this case to be bell-shaped
rather than being a step function. Surprisingly, $\Delta(0)$ is
however in much closer agreement with its approximation, as shown on
Fig.~%
\ref{fig:WedApr7114819BST2010}(b) and~(c) in the weak and strong
coupling regime, respectively. There are small quantitative deviations
in~(b) between the numerical points and the formula when using the
exact parameters of the potential. By fitting the numerical results, a
perfect agreement can be found for slight variations of
$\hbar\omega_\mathrm{D}$ and $U_0$. The BCS approximation therefore
turns out to be an exceedingly good one as compared to an exact
solution of the gap equation. For the procedure to make sense, similar
results should be obtained for a smoothed well that approximates the BCS
square well~\cite{ginzburg82a}. We will not address this point here,
but go directly to the case where the potential is not always
attractive.

\subsection{Bogoliubov potential}

\begin{figure}[tbp]
\centering
\includegraphics[width=.65\linewidth]{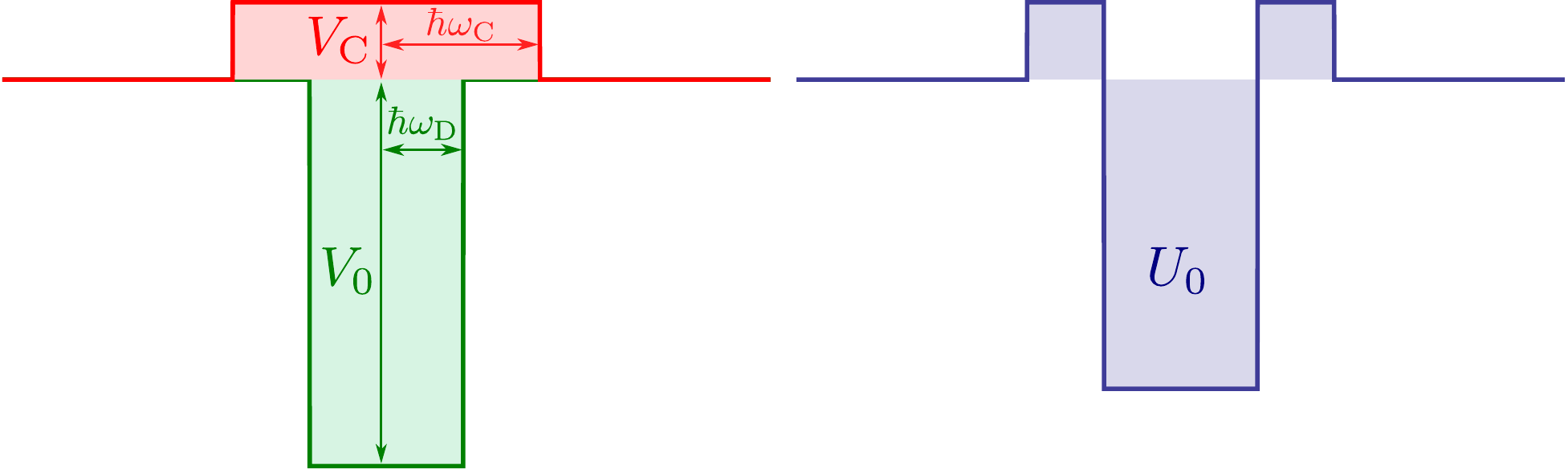}
\caption{Adding an overall repulsive Coulomb repulsion $V_\mathrm{C}$ (red,
left) until a cutoff $\protect\omega_\mathrm{C}$ to the BCS potential $V_0$
(green, left) results in the Bogoliubov stepwise potential (right). We take
the convention $V_0$, $V_\mathrm{C}$, $U_0$ positive in the above
representation and $U_0=V_0-V_\mathrm{C}$, so that, e.g., $V_\mathrm{C}<0$
means attractive contribution of the ``Coulomb repulsion''.}
\label{fig:WedApr7153152BST2010}
\end{figure}

The potential is not always attractive when, for instance, some overall
repulsion, such as direct Coulomb interaction, is superimposed on the
attractive Cooper potential, as shown on Fig.~\ref{fig:WedApr7153152BST2010}%
. The Coulomb interaction is time independent and should therefore extend to
all $\omega$ but here also a cutoff $\omega_\mathrm{C}$ is introduced to
avoid divergencies. This results in an attractive, Cooper-like potential,
flanked by two repulsive windows. Such a potential is known as the Bogoliubov
potential~\cite{ketterson_book99a,zheng05a}.

This approximation has been used to show extremely counter-intuitive
behaviour of the gap equation and justify a-posteriori another
heavily criticized approximation of BCS, neglecting Coulomb repulsion:
the BCS mechanism indeed assumes only attraction between electrons,
which can be dimmed by Coulomb repulsion, but which never explicitly
appears as such (like in the Bogoliubov scenario). The great result of
Cooper was that binding occurs at arbitrarily small attraction. An
important result of the Bogoliubov potential is to show that the
detrimental effects of Coulomb repulsion are greatly reduced in the
gap~\cite{zheng05a}.

We now present a linearization of the gap equation that allows one to obtain an
approximate solution for the critical temperature~\cite{ketterson_book99a}.
By assuming the gap equation to be a two step valued function~$\Delta
=(\Delta _{1},\Delta _{2})^{T}$, the gap equation becomes $(I-1)\Delta =0$
with 
\begin{equation}
I=-%
\begin{pmatrix}
-U_{0}\mathcal{I}_{1} & V_{\mathrm{C}}\mathcal{I}_{2} \\ 
V_{\mathrm{C}}\mathcal{I}_{1} & V_{\mathrm{C}}\mathcal{I}_{2}%
\end{pmatrix}%
\,,  \label{eq:TueMay11120248BST2010}
\end{equation}%
where
\begin{subequations}
\label{eq:WedMay26103213BST2010}
\begin{align}
\mathcal{I}_{1}& =\int_{-\hbar \omega _{\mathrm{D}}}^{\hbar \omega _{\mathrm{%
D}}}\frac{\tanh (\xi /(2k_{\mathrm{B}}T_{\mathrm{C}}))}{\sqrt{\xi
^{2}+\Delta _{1}^{2}}}d\xi \,,\approx \ln \left( \frac{1.13\hbar \omega _{%
\mathrm{D}}}{k_{\mathrm{B}}T_{\mathrm{C}}}\right) \,,
\label{eq:WedMay26103747BST2010} \\
\mathcal{I}_{2}& =\int_{\hbar \omega _{\mathrm{D}}}^{\hbar \omega _{\mathrm{C%
}}}\frac{\tanh (\xi /(2k_{\mathrm{B}}T_{\mathrm{C}}))}{\sqrt{\xi ^{2}+\Delta
_{2}^{2}}}d\xi \approx \ln \left( \frac{\omega _{\mathrm{C}}}{\omega _{%
\mathrm{D}}}\right) \,,  \label{eq:WedMay26103812BST2010} 
\end{align}
appear invariably in the columns of $I$. In Eq.~(\ref%
{eq:WedMay26103747BST2010}), the BCS approximation has been applied while in
Eq.~(\ref{eq:WedMay26103812BST2010}), the fact that $|\xi |\gg 0$ has been
used to neglect $\Delta $ in the denominator and the temperature in the
numerator. The parametrization of the matrix in terms of the potential
depends on the particular configuration (for example the relative widths 
of the various layers of the structure). Here we have adopted the
original parametrization of Bogoliubov, which assumes narrow repulsive
elbows surrounding a large attractive central region. Solving the linear
equation for $\mathcal{I}_{1}$ and then for the critical temperature, we
find: 
\end{subequations}
\begin{equation}
k_{\mathrm{B}}T_{\mathrm{C}}\approx 1.13\hbar \omega _{\mathrm{D}}\exp
\left( -\frac{1}{V_{0}-\displaystyle\frac{V_{\mathrm{C}}}{1+\mathcal{I}%
_{2}V_{\mathrm{C}}}}\right) \,.  \label{eq:WedMay26105415BST2010}
\end{equation}

In this form, one can see how Coulomb repulsion indeed introduces a small
correction to the original BCS formula. The expression also seems to
indicate that $U_0$ could be repulsive ($V_0-V_\mathrm{C}<0$) and
still lead to a gap as long as the denominator in
Eq.~(\ref{eq:WedMay26105415BST2010}) remains positive.

In the following, we compare these predictions with numerical solutions of
the gap equation, keeping in mind that the iterative procedure is not
assured, mathematically, to converge. We have observed that indeed, it
sometimes encounters problems and exhibit strong instabilities, with
bifurcations of solutions, for example.

\begin{figure}[tbp]
\centering
\includegraphics[width=\linewidth]{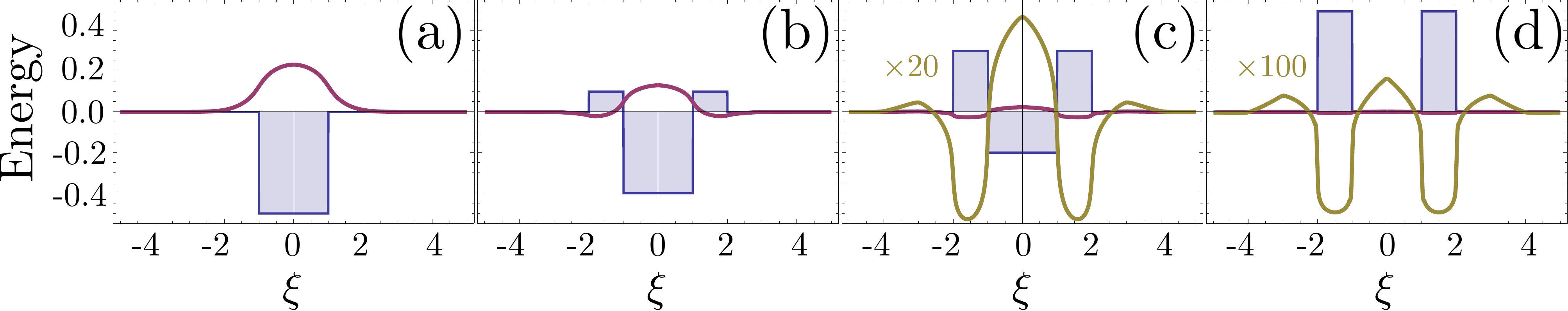}
\caption{Gap (thick magenta and, magnified, thick khaki) of the Bogoliubov
potential (filled blue) solved numerically for the parameters: $\hbar%
\protect\omega_\mathrm{D}=1$, $\hbar\protect\omega_\mathrm{C}=2$, $V_0=1$
and $V_\mathrm{C}$ taking values from (a) to~(d) of: 0 (BCS), 0.1, 0.3 and
0.495. Coulomb repulsion results in a dip in the gap function that, for
increasing values, results in oscillations in the gap function.
Paradoxically, even when it is large and dominating attraction, repulsion
does not prevent a gap, as seen in (d).}
\label{fig:ThuApr8100431BST2010}
\end{figure}

In Fig.~\ref{fig:ThuApr8100431BST2010}, we show the evolution of the gap
function as $V_\mathrm{C}$ is increased, from zero (BCS) to a point where
the overall potential is essentially repulsive. With onset of the repulsion,
the gap acquires two negative sides and becomes a
highly distorted function for large $V_\mathrm{C}$. Note that the
approximation of constant gap over the various regions is at least as good
as for the case of BCS.

\begin{figure}[tbp]
\centering
\includegraphics[width=\linewidth]{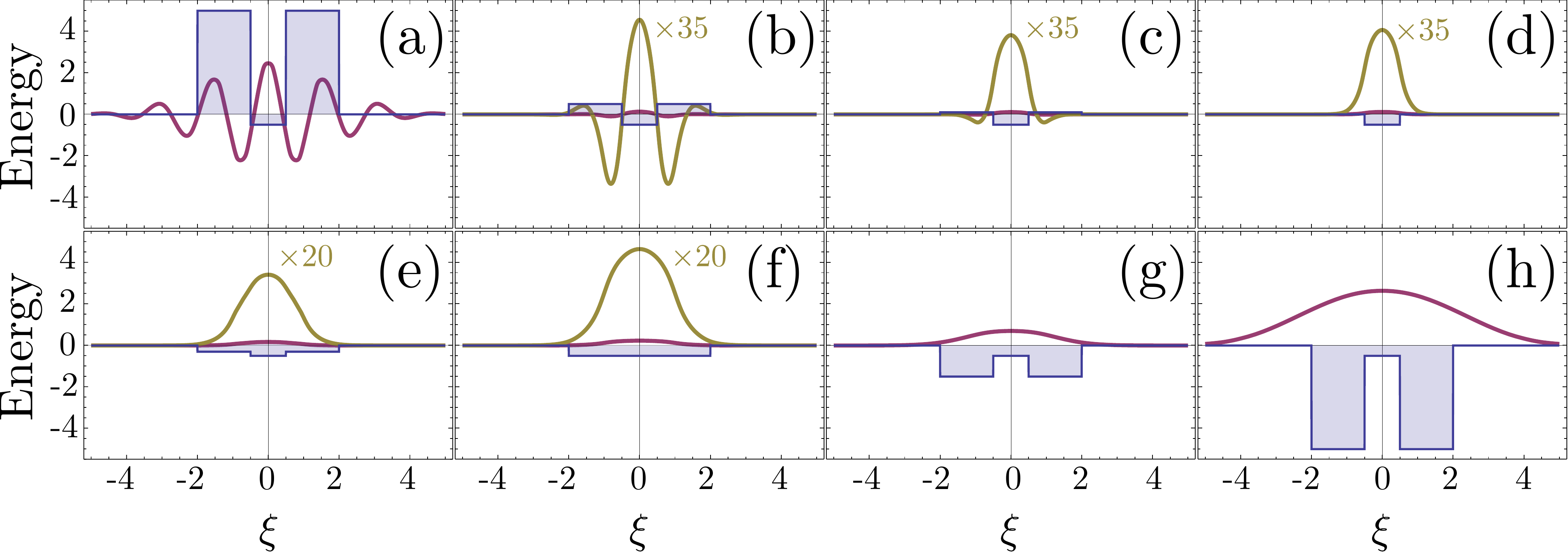}
\caption{Gap (thick magenta and, magnified, thick khaki) of the Bogoliubov
potential (filled blue) solved numerically for the following parameters: $\hbar\protect%
\omega_\mathrm{D}=0.5$, $\hbar\protect\omega_\mathrm{C}=2$, $V_0-V_\mathrm{C}%
=0.5$ (fixed) and $V_\mathrm{C}$ (the height of the elbow) taking values
from (a) to~(h) of: -5, -0.5, -0.1, 0 (BCS), 0.3, 0.5 (BCS), 1.5 and 5. The
repulsive nature of the elbows change the character of the gap from
bell-shaped to an oscillating function. The oscillating gap in the presence of
strong repulsion allows, paradoxically, high values of the gap at
zero-exchanged energy.}
\label{fig:WedApr7205340BST2010}
\end{figure}

In Fig.~\ref{fig:WedApr7205340BST2010}, we now show the case where $U_0=V_0-V_\mathrm{C}$ is
held constant as the strength of the repulsion is varied independently. In
foresight of what is to come later, we also allow the elbows to be negative,
that is, to contribute an additional attraction to the conventional
mechanism (for now we do not consider physical justification of this). Another
unexpected result is obtained: the repulsive potential is in this case
favouring a larger gap, as can be seen by comparing (a), where the gap function is
highly oscillatory, to (d) where it recovers the BCS bell-shape. In (h),
the distorted but overall attractive potential still results in a BCS type
gap, but wider and larger. Note how the repulsion, by ``squeezing''
the gap, allows it to achieve much higher values than for the case of 
smaller or no repulsion.

These unexpected results are confirmed phenomenologically by the Bogoliubov
approximation Eq.~(\ref{eq:WedMay26105415BST2010}), which we plot as a dashed
line in Fig.~\ref{fig:WedApr7205340BST2010}, compared to $\Delta (0)$
computed numerically. Here we should emphasize that the two quantities are
not meant to be compared quantitatively (as we did when comparing the BCS
formula with the numerical solution), since one, $\Delta (0)$ (computed
numerically) is the gap at zero temperature while the other, $k_{\mathrm{B}%
}T_{\mathrm{C}}$ is the temperature at which the gap vanishes. There is a
monotonous relationship between the two, that is, increasing $\Delta (0)$
implies increasing $T_{\mathrm{C}}$, so one can appreciate the consistency of
the results by observing similar trends. The obstacle to conducting an
extensive numerical comparison is that it is an intensive task numerically to
compute $T_{\mathrm{C}}$, since this requires solution of the gap equation for
various temperatures until the curve $\Delta (T)$ is obtained and its
intersect with zero is found. A critical slowing down phenomenon
makes the iterative process slower as the critical temperature is
approached. In addition, numerical instabilities are stronger at nonzero $T$. Therefore, although it is relatively straightforward to compute $%
\Delta (0)$ numerically, it is not convenient to use this method to obtain $%
T_{\mathrm{C}}$. On the other hand, the Bogoliubov approximation gives a
fair estimate of $T_{\mathrm{C}}$, but is not able to provide the gap at zero
temperature, since at the core of its method, there is an assumption of
vanishing $\Delta $. Therefore, we have two complementary methods, each
suited to provide a relevant aspect of the problem. We note that the gap
at zero temperature is an important quantity which can be measured
independently from $T_{\mathrm{C}}$ by Andreev reflection in conductivity
experiments, which is why we discuss it here in a great detail. 

We can see, indeed, that the qualitative agreement is good and that the
counter-intuitive features of the gap equation are reproduced by the
analytical formula.

\begin{figure}[tbp]
\centering
\includegraphics[width=\linewidth]{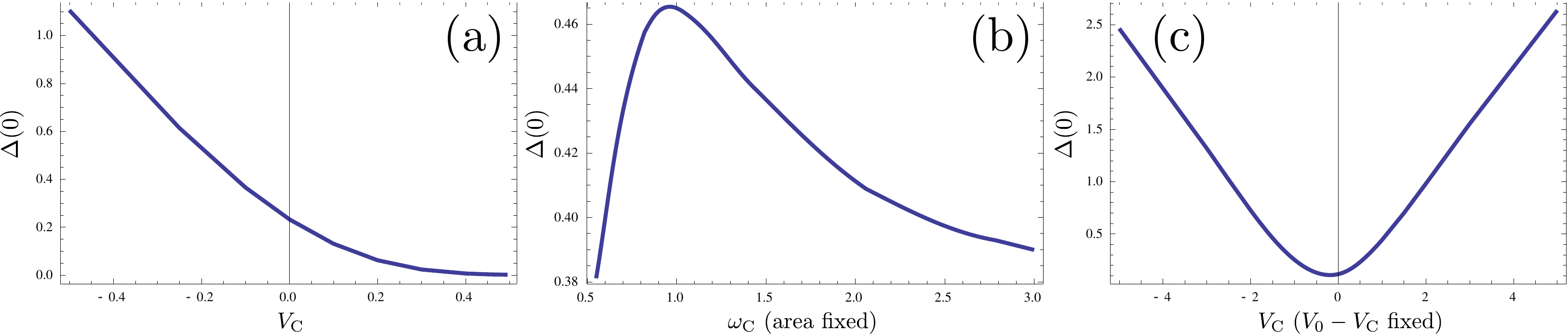}
\caption{$\Delta(0)$ for the case of Fig.~\protect\ref%
{fig:ThuApr8100431BST2010} (a) and \protect\ref{fig:WedApr7205340BST2010}
(c), for their respective parameters, and (b), with $\hbar\protect\omega_%
\mathrm{D}=.5$, $V_0=1$, $\protect\omega_\mathrm{C}$ changing as indicated
and $V_\mathrm{C}$ changing such that the area of the repulsive elbow is
conserved.}
\label{fig:ThuApr8102721BST2010}
\end{figure}


\subsection{Polariton potential}

In the case of the polariton problem, we have seen that, even when
neglecting Coulomb repulsion (as in the original BCS formulation), the
potential $U(\omega)$ departs strongly from the Cooper potential and
features two large attractive regions far from small energies, immediately
followed by two strong repulsive windows. We extend the Bogoliubov method to
a three-step approximation of this potential, such as displayed in Fig.~\ref%
{fig:SatApr10115804BST2010}, with, in reference to previous potentials,
notations $\omega_\mathrm{D}$, $\omega_\mathrm{C}$ and $\omega_\mathrm{B}$
for the boundaries of the central, shallow attractive region, narrow, deep
attractive region and repulsive region, respectively. This is a notation
only and is not mean to be understood as referring to Debye, Coulomb or
Bogoliubov in any strict sense. Following the same premises, we approximate
the gap equation by a three-step valued function~$\Delta=(\Delta_1,
\Delta_2, \Delta_3)^T$. This approximation turns out to be an exceedingly
good one in certain cases, such as the one displayed in Fig.~\ref%
{fig:SatApr10115804BST2010}. Here there is even more room to choose a
parametrization of the $I$ matrix. We now give general guidelines on how to
build this matrix. The simplest method is to fix $\xi$ on the lhs of Eq.~(%
\ref{eq:MonMar29151237BST2010}) at the center of each region and, in the
corresponding row, take for each column the potential that is sampled more
by the difference $|\xi-\xi^{\prime }|$. Refinements are possible, such as
weighting elements of $I$ by coefficients which reflect how much time the
variables $\xi$ and $\xi^{\prime }$ spend in the regions that determine the
matrix equation. This problem has the following mathematical expression:
how is the random variable $X-Y$ distributed when $X$ (resp.~$Y$) is
uniformly distributed in an interval $[g_i, g_{i+1}]$ (resp.~$[h_i, h_{i+1}]$%
). The solution is easily obtained as proportional (normalize to unity) to: 
\begin{equation}  \label{eq:WedMay19161456BST2010}
P(X-Y=\theta)\propto\sqrt{[\max(g_i,h_i-\theta)]-\min(g_{i+1},h_{i+1}-%
\theta)]^2\newline
+[\max(\theta+g_i,h_i)-\min(\theta+g_{i+1},h_{i+1}]^2}\,.
\end{equation}
This is easily obtained geometrically (the square root comes from
Pythagoras' theorem) and the problem results in finding the intersect
of a line with the grid. There are two configurations. Working out the
cases shows that Eq.~(%
\ref{eq:WedMay19161456BST2010}) reduces to a triangular or a top-head
truncated triangular distribution. The coefficients entering $I$ can
then be taken as the potentials weighted by the area intersecting
their corresponding region. The soundness of such an approach can be
appreciated only if cases are checked numerically to compare
quantitatively various parametrizations. For simplicity, we shall here
consider cases where only the dominant potential is considered. An
example of such a gap equation $%
(I-1)\Delta=0$ is defined with:
\begin{equation}  \label{eq:FriDec4114730GMT2009}
I=- 
\begin{pmatrix}
V_0\mathcal{I}_1 & V_1\mathcal{I}_2 & -V_\mathrm{C}\mathcal{I}_3 \\ 
V_1\mathcal{I}_1 & V_0\mathcal{I}_2 & V_0\mathcal{I}_3 \\ 
-V_\mathrm{C}\mathcal{I}_1 & V_0\mathcal{I}_2 & V_0\mathcal{I}_3%
\end{pmatrix}%
\,,
\end{equation}
where 
\begin{equation}  \label{eq:FriDec4115035GMT2009}
\mathcal{I}_i=\int_{\hbar\omega_i^<}^{\hbar\omega_i^>}\frac{\tanh(\xi/(2k_%
\mathrm{B}T_\mathrm{C}))}{\sqrt{\xi^2+\Delta_i^2}}d\xi\,,
\end{equation}
and $\mathcal{I}_i$ is integrated on the respective steps, as defined by
the integral boundary conditions, i.e., $(\omega_1^<,\omega_1^>,\omega_2^<,%
\omega_2^>,\omega_3^<,\omega_3^>)=(0,\omega_\mathrm{D},\omega_\mathrm{D}%
,\omega_\mathrm{C},\omega_\mathrm{C},\omega_\mathrm{B})$.

\begin{figure}[tbp]
\centering
\includegraphics[width=.5\linewidth]{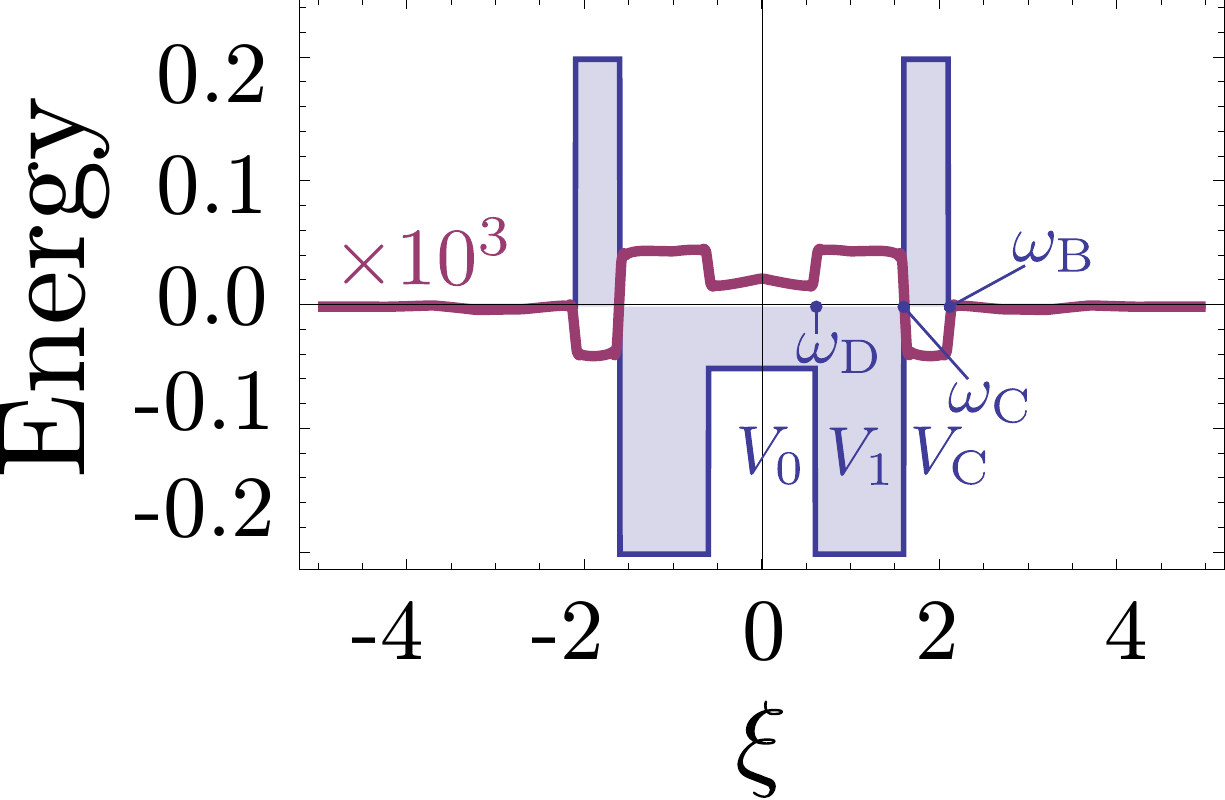}
\caption{(solid blue) Three-step potential that models the form of
effective electron-electron interaction when polariton-mediated with the
three parameters $\protect\omega_{\mathrm{B,C,D}}$ and, (thick purple) the
gap $\Delta(\protect\xi)$ as computed numerically for this potential. In
good approximation, the gap is also itself a three-steps function.}
\label{fig:SatApr10115804BST2010}
\end{figure}

As before, we estimate the critical temperature by the condition that gives
vanishing values of the gap on the whole interval. The first integral $%
\mathcal{I}_1$ is evaluted in the same approximation as for the usual BCS
(or Bogoliubov) potential: 
\begin{equation}  \label{eq:SatApr10121736BST2010}
\mathcal{I}_1=\ln\left(1.13\frac{\hbar\omega_\mathrm{D}}{k_\mathrm{B}T_%
\mathrm{C}}\right)\,.
\end{equation}

$\mathcal{I}_2$ and $\mathcal{I}_3$ are also essentially logarithmic in
their energy range and we take: 
\begin{equation}  \label{eq:SatApr10121545BST2010}
\mathcal{I}_2=\ln(\omega_\mathrm{C}/\omega_\mathrm{D})\,,\quad\text{and}%
\quad \mathcal{I}_3=\ln(\omega_\mathrm{B}/\omega_\mathrm{C})\,.
\end{equation}

We solve the gap equation by setting its determinant to zero and solving for 
$T_\mathrm{C}$, giving %

\begin{equation}  \label{eq:SunApr11141817BST2010}
k_\mathrm{B}T_\mathrm{C}=1.13\hbar\omega_\mathrm{D}\exp\left(- \frac{1}{V_0} 
\frac{1} {1+ \frac{ \ln\left(\frac{\omega_\mathrm{C}}{\omega_\mathrm{D}}%
\right) \ln\left(\frac{\omega_\mathrm{B}}{\omega_\mathrm{C}}\right) (V_1+V_%
\mathrm{C})^2 +\ln\left(\frac{\omega_\mathrm{C}}{\omega_\mathrm{D}}\right)%
\frac{V_1^2}{V_0} +\ln\left(\frac{\omega_\mathrm{B}}{\omega_\mathrm{C}}%
\right)\frac{V_\mathrm{C}^2}{V_0}} {1+\ln\left(\frac{\omega_\mathrm{B}}{%
\omega_\mathrm{D}}\right)V_0} } \right)\,.
\end{equation}

\begin{figure}[tbp]
\centering
\includegraphics[width=.5\linewidth]{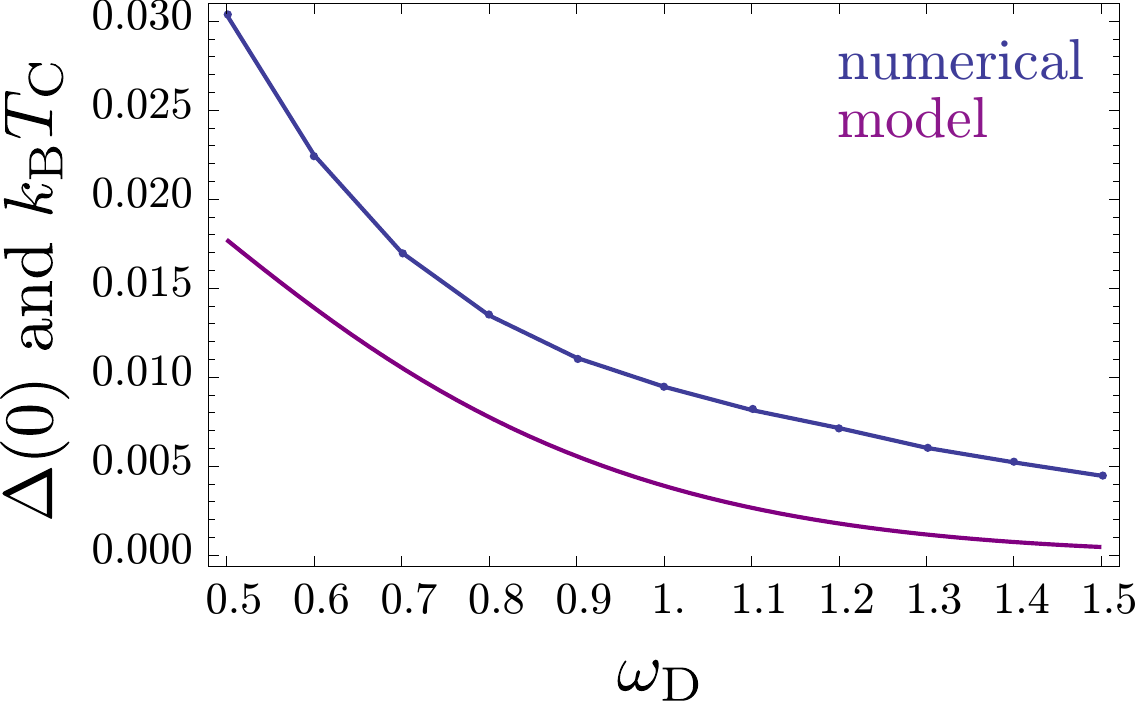}
\caption{$\Delta(0)$ (joined points, blue) and $k_\mathrm{B}T_\mathrm{C}$
from the three-step Bogoliubov approximation, in the case where the width
of the central attraction is increased. Unexpectedly, this results in a
decrease of the critical temperature.}
\label{fig:SatMay8175257BST2010}
\end{figure}

We see that the impact of the polariton potential shape on the gap is rather
intricate, here as well, with some unexpected behaviour, produced
both numerically and from this formula. In Fig.~\ref%
{fig:SatMay8174949BST2010}, we show the effect of widening the central,
attractive region of the potential. Naively one would expect this to
increase the gap (or critical temperature), since $\Delta(0)$ increases
linearly with $\hbar\omega_\mathrm{D}$. However, Eq.~(\ref%
{eq:SunApr11141817BST2010}) predicts a decrease of the critical temperature,
as shown in Fig.~\ref{fig:SatMay8175257BST2010}. The reason why is
understood by computing the gap at zero temperature, as seen in Fig.~\ref%
{fig:SatMay8174949BST2010}, where the repulsive barrier is shown to behave
as a trap for the gap function, which results, as the width is increased, in
a loosening of the gap strength, akin to the quantum mechanical situation of
a bound state in a square potential. The numerical and linearized models
display a good qualitative agreement, as seen in Fig.~\ref%
{fig:SatMay8175257BST2010}. Let us repeat that we are not considering here
quantities that can be directly compared, since $\Delta(T_\mathrm{C})$
should be used for this purpose.

\begin{figure}[tbp]
\centering
\includegraphics[width=\linewidth]{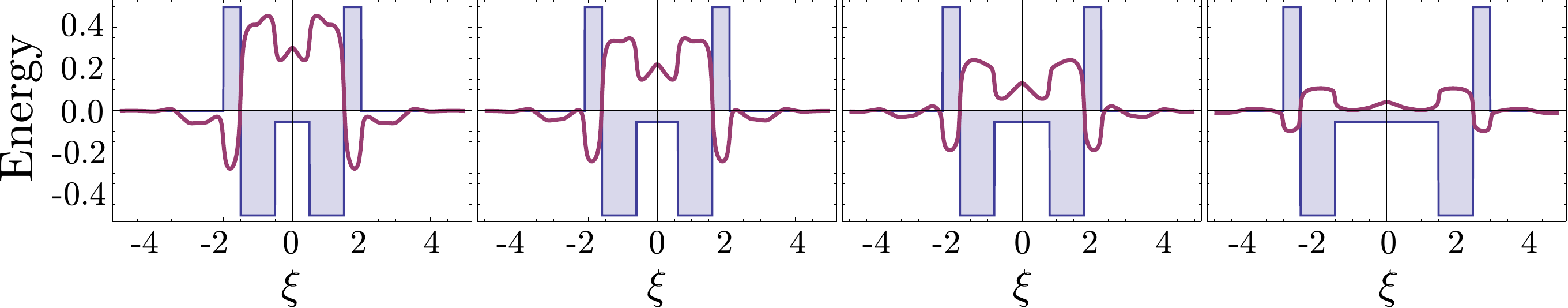}
\caption{Potential (filled blue) and its corresponding gap equation, solved
numerically, as the size of the central attractive plateau is increased.
Somewhat unexpectedly, the gap decreases as a result. The repulsive barriers
have the effect of squeezing the gap up.}
\label{fig:SatMay8174949BST2010}
\end{figure}

\begin{figure}[tbp]
\centering
\includegraphics[width=\linewidth]{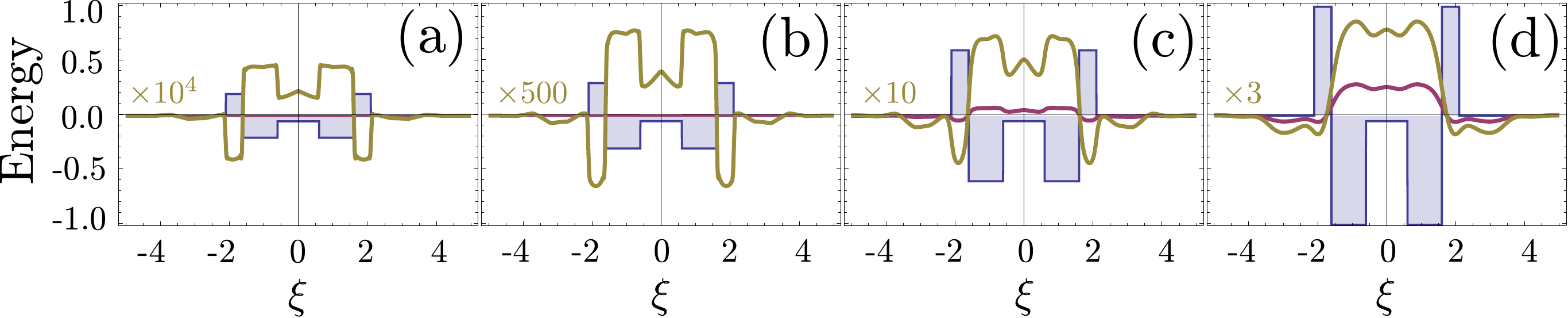}
\caption{Numerical solution of the gap equation when both the attractive and
repulsive parts of the potential are increased together (here in the same proportions), showing a rapid increase of the gap.}
\label{fig:SatMay8181120BST2010}
\end{figure}

\begin{figure}[tbp]
\centering
\includegraphics[width=\linewidth]{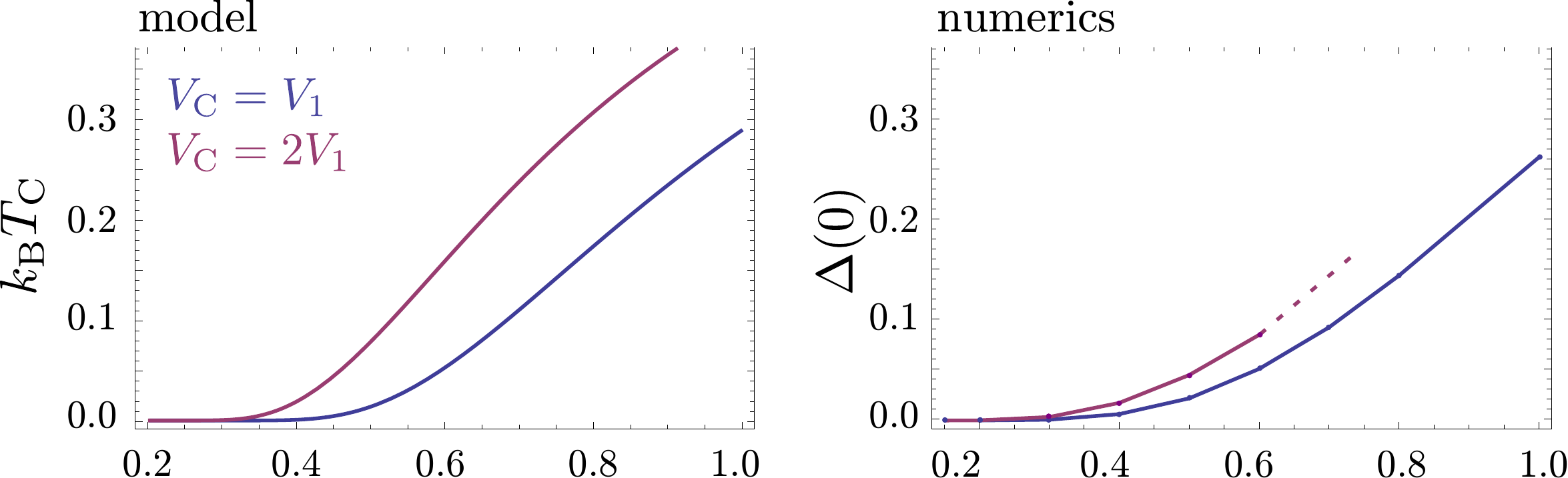}
\caption{$\Delta(0)$ (joined points, blue) and $k_\mathrm{B}T_\mathrm{C}$
from the three-step Bogoliubov approximation.}
\label{fig:SatMay8181123BST2010}
\end{figure}

Another example is shown in Fig.~\ref{fig:SatMay8181120BST2010}, where both
the attractive and repulsive parts of the potential are increased together.
This results in a rapid increase of the gap. The result is also confirmed by
numerical simulations, as shown in Fig.~\ref{fig:SatMay8181123BST2010}, and
a good agreement is obtained. Unexpectedly,
if the repulsive part grows twice as fast as the attractive one, not only
does this also result in an increase of the gap, but even a faster one.
This, too, is confirmed by the numerics. Note that in this case, we reach
a region where our numerical procedure jumps to other solutions (usually of
a highly oscillatory character, such as those reported in~\cite{zheng05a}),
but behaves as expected until then. In this regard, the value of having an
approximate analytical solution is obvious.

\begin{figure}[tbp]
\centering
\includegraphics[width=.5\linewidth]{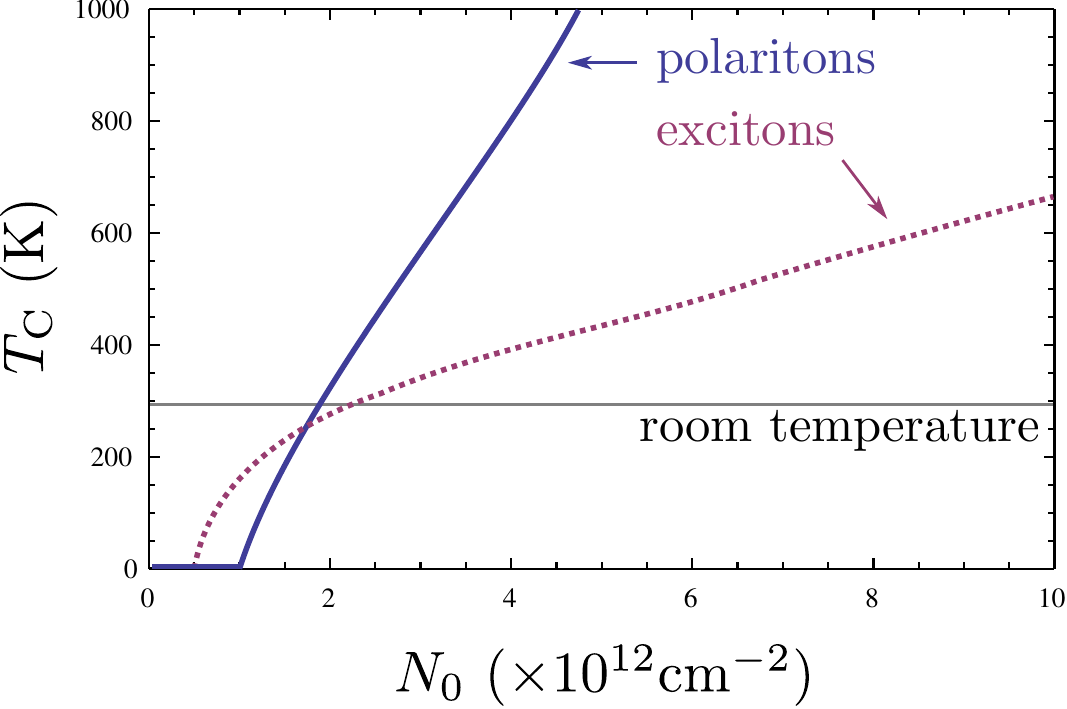}
\caption{(colour online) Critical temperature as a function of the
condensate density~$N_0$ for the case of a polariton condensate (solid blue)
and of an exciton condensate (dotted purple).}
\label{fig:SatAug21082849BST2010}
\end{figure}

This analysis is finally applied to the case of the gap equation with
the numerical results of the critical temperature obtained in both the
polariton and exciton cases for the cases of
Figs.~\ref{fig:SatMay29135948BST2010} and
\ref{fig:SatAug21093037BST2010}. The parameters are gathered in
Table~\ref{tab:MonNov29002411CET2010} and the results plotted in
Fig.~\ref{fig:SatAug21082849BST2010}. Both cases show a strong
variation of the critical temperature with moderate variations of the
condensate density (one order of magnitude). The polariton case is
steeper and roughly linear while the exciton case increases less
quickly but begins earlier.  In both cases, temperatures are very
high, and other effects will surely break the mechanism, for instance
the loss of the condensate. This shows, however, the robustness of the
mechanism in the conditions where it should apply.

\section{Conclusions}

\label{sec:TueMay25182435BST2010}

We have studied possible mechanisms of superconductivity in
semiconductor heterostructure systems where a Bose-Einstein condensate
mediating the effective electron-electron interaction leads to
enhancement of the coupling, yielding very high critical
temperatures. We have considered the case of an exciton BEC,
consisting of a sandwich of an $n$-doped QW containing the
superconducting electrons, in contact with coupled QWs where the BEC
of indirect excitons is formed, for instance by optical excitation. We
have also considered the case of a polariton BEC, where $n$-doped
QWs and undoped QWs hosting excitons are embedded in a microcavity. We
have computed in these two cases the effective electron-electron
interaction~$U(\omega )$ as a function of the exchanged energy~$\hbar
\omega $, showing how the retardation effect acquires a peculiar
character in the polariton case, namely, yielding a weakly attractive
potential at long times, followed by a succession of strongly
attractive and strongly repulsive windows. The gap equation in this
case exhibits strong differences as opposed to the case of the Cooper
potential (of conventional BCS and exciton-BEC mechanism). To
understand the physical mechanism leading to large gaps and bypass
numerical instabilities, we studied in detail the gap equation in the
cases of simplified step-wise potentials, and offered an analytical
method to obtain the critical temperature, which is in qualitative
agreement with the numerical results. Our results suggest record
breaking critical temperatures in these systems. We stress, however,
that the path towards achievement of exciton mediated
superconductivity at high temperatures may be long and full of
obstacles. Of the two experimentally relevant systems that we have
considered, one (coupled QWs)\ only shows the BEC of excitons at very
low temperatures (less than 1K~\cite{butov02a}). Consequently, one
cannot expect high temperature superconductivity in this system, while
at low temperatures the superconducting gap may be very large. In
microcavities, polariton BEC\ or polariton lasing have indeed been
demonstrated at room temperature \cite%
{christopoulos07a,baumberg08a}. However, embedding a high-quality
$n$-doped QW\ inside the cavity, the proper choice of the experimental
geometry in order to minimise optical absorption in the doped QW, and
especially fabrication of quantum contacts for selective injection of
carriers in the QW of interest, may pose technological
difficulties. It is possible that hybrid metal-semiconductor or semi-metal
semiconductor systems may appear more suitable for the observation of
the predicted effects. A conventional superconductor put in contact
with a semiconductor also seems promising.  Collective quantum
phenomena in Bose-Fermi mixtures are extremely complicated and we
foresee breakthroughs in their study in multilayer structures
combining a Fermi gas of electrons and an exciton BEC.

\begin{acknowledgements}
  We are grateful to H. Ouerdanne and P. Lagoudakis for useful
  discussions. Support from EPSRC and EU FP7 ITN project CLERMONT4 are
  acknowledged. I.A.S. thanks the support from Rannis ``Center of
  excellence in polaritonics'' and FP7 IRSES POLAPHEN project.
\end{acknowledgements}

\bibliography{books,Sci,arXiv,superconductivity-polaritons}

\begin{thebibliography}{55}%
\makeatletter
\providecommand \@ifxundefined [1]{%
 \@ifx{#1\undefined}
}%
\providecommand \@ifnum [1]{%
 \ifnum #1\expandafter \@firstoftwo
 \else \expandafter \@secondoftwo
 \fi
}%
\providecommand \@ifx [1]{%
 \ifx #1\expandafter \@firstoftwo
 \else \expandafter \@secondoftwo
 \fi
}%
\providecommand \natexlab [1]{#1}%
\providecommand \enquote  [1]{``#1''}%
\providecommand \bibnamefont  [1]{#1}%
\providecommand \bibfnamefont [1]{#1}%
\providecommand \citenamefont [1]{#1}%
\providecommand \href@noop [0]{\@secondoftwo}%
\providecommand \href [0]{\begingroup \@sanitize@url \@href}%
\providecommand \@href[1]{\@@startlink{#1}\@@href}%
\providecommand \@@href[1]{\endgroup#1\@@endlink}%
\providecommand \@sanitize@url [0]{\catcode `\\12\catcode `\$12\catcode
  `\&12\catcode `\#12\catcode `\^12\catcode `\_12\catcode `\%12\relax}%
\providecommand \@@startlink[1]{}%
\providecommand \@@endlink[0]{}%
\providecommand \url  [0]{\begingroup\@sanitize@url \@url }%
\providecommand \@url [1]{\endgroup\@href {#1}{\urlprefix }}%
\providecommand \urlprefix  [0]{URL }%
\providecommand \Eprint [0]{\href }%
\providecommand \doibase [0]{http://dx.doi.org/}%
\providecommand \selectlanguage [0]{\@gobble}%
\providecommand \bibinfo  [0]{\@secondoftwo}%
\providecommand \bibfield  [0]{\@secondoftwo}%
\providecommand \translation [1]{[#1]}%
\providecommand \BibitemOpen [0]{}%
\providecommand \bibitemStop [0]{}%
\providecommand \bibitemNoStop [0]{.\EOS\space}%
\providecommand \EOS [0]{\spacefactor3000\relax}%
\providecommand \BibitemShut  [1]{\csname bibitem#1\endcsname}%
\let\auto@bib@innerbib\@empty
\bibitem [{\citenamefont {Kavokin}\ \emph
  {et~al.}(2007{\natexlab{a}})\citenamefont {Kavokin}, \citenamefont
  {Baumberg}, \citenamefont {Malpuech},\ and\ \citenamefont
  {Laussy}}]{kavokin_book07a}%
  \BibitemOpen
  \bibfield  {author} {\bibinfo {author} {\bibfnamefont {A.}~\bibnamefont
  {Kavokin}}, \bibinfo {author} {\bibfnamefont {J.~J.}\ \bibnamefont
  {Baumberg}}, \bibinfo {author} {\bibfnamefont {G.}~\bibnamefont {Malpuech}},
  \ and\ \bibinfo {author} {\bibfnamefont {F.~P.}\ \bibnamefont {Laussy}},\
  }\href@noop {} {\emph {\bibinfo {title} {Microcavities}}}\ (\bibinfo
  {publisher} {Oxford University Press},\ \bibinfo {year} {2007})\BibitemShut
  {NoStop}%
\bibitem [{\citenamefont {Deng}\ \emph {et~al.}(2010)\citenamefont {Deng},
  \citenamefont {Haug},\ and\ \citenamefont {Yamamoto}}]{deng10a}%
  \BibitemOpen
  \bibfield  {author} {\bibinfo {author} {\bibfnamefont {H.}~\bibnamefont
  {Deng}}, \bibinfo {author} {\bibfnamefont {H.}~\bibnamefont {Haug}}, \ and\
  \bibinfo {author} {\bibfnamefont {Y.}~\bibnamefont {Yamamoto}},\ }\href@noop
  {} {\bibfield  {journal} {\bibinfo  {journal} {Rev. Mod. Phys.}\ }\textbf
  {\bibinfo {volume} {82}},\ \bibinfo {pages} {1489} (\bibinfo {year}
  {2010})}\BibitemShut {NoStop}%
\bibitem [{\citenamefont {{\u Imamo\=glu}}\ \emph {et~al.}(1996)\citenamefont
  {{\u Imamo\=glu}}, \citenamefont {Ram}, \citenamefont {Pau},\ and\
  \citenamefont {Yamamoto}}]{imamoglu96a}%
  \BibitemOpen
  \bibfield  {author} {\bibinfo {author} {\bibfnamefont {A.}~\bibnamefont {{\u
  Imamo\=glu}}}, \bibinfo {author} {\bibfnamefont {R.~J.}\ \bibnamefont {Ram}},
  \bibinfo {author} {\bibfnamefont {S.}~\bibnamefont {Pau}}, \ and\ \bibinfo
  {author} {\bibfnamefont {Y.}~\bibnamefont {Yamamoto}},\ }\href@noop {}
  {\bibfield  {journal} {\bibinfo  {journal} {Phys. Rev. A}\ }\textbf {\bibinfo
  {volume} {53}},\ \bibinfo {pages} {4250} (\bibinfo {year}
  {1996})}\BibitemShut {NoStop}%
\bibitem [{\citenamefont {Baumberg}\ \emph {et~al.}(2000)\citenamefont
  {Baumberg}, \citenamefont {Savvidis}, \citenamefont {Stevenson},
  \citenamefont {Tartakovskii}, \citenamefont {Skolnick}, \citenamefont
  {Whittaker},\ and\ \citenamefont {Roberts}}]{baumberg00a}%
  \BibitemOpen
  \bibfield  {author} {\bibinfo {author} {\bibfnamefont {J.~J.}\ \bibnamefont
  {Baumberg}}, \bibinfo {author} {\bibfnamefont {P.~G.}\ \bibnamefont
  {Savvidis}}, \bibinfo {author} {\bibfnamefont {R.~M.}\ \bibnamefont
  {Stevenson}}, \bibinfo {author} {\bibfnamefont {A.~I.}\ \bibnamefont
  {Tartakovskii}}, \bibinfo {author} {\bibfnamefont {M.~S.}\ \bibnamefont
  {Skolnick}}, \bibinfo {author} {\bibfnamefont {D.~M.}\ \bibnamefont
  {Whittaker}}, \ and\ \bibinfo {author} {\bibfnamefont {J.~S.}\ \bibnamefont
  {Roberts}},\ }\href@noop {} {\bibfield  {journal} {\bibinfo  {journal} {Phys.
  Rev. B}\ }\textbf {\bibinfo {volume} {62}},\ \bibinfo {pages} {R16247}
  (\bibinfo {year} {2000})}\BibitemShut {NoStop}%
\bibitem [{\citenamefont {Skolnick}\ \emph {et~al.}(1998)\citenamefont
  {Skolnick}, \citenamefont {Fisher},\ and\ \citenamefont
  {Whittaker}}]{skolnick98a}%
  \BibitemOpen
  \bibfield  {author} {\bibinfo {author} {\bibfnamefont {M.~S.}\ \bibnamefont
  {Skolnick}}, \bibinfo {author} {\bibfnamefont {T.~A.}\ \bibnamefont
  {Fisher}}, \ and\ \bibinfo {author} {\bibfnamefont {D.~M.}\ \bibnamefont
  {Whittaker}},\ }\href@noop {} {\bibfield  {journal} {\bibinfo  {journal}
  {Semicond. Sci. Technol.}\ }\textbf {\bibinfo {volume} {13}},\ \bibinfo
  {pages} {645} (\bibinfo {year} {1998})}\BibitemShut {NoStop}%
\bibitem [{\citenamefont {Savvidis}\ \emph {et~al.}(2000)\citenamefont
  {Savvidis}, \citenamefont {Baumberg}, \citenamefont {Stevenson},
  \citenamefont {Skolnick}, \citenamefont {Whittaker},\ and\ \citenamefont
  {Roberts}}]{savvidis00a}%
  \BibitemOpen
  \bibfield  {author} {\bibinfo {author} {\bibfnamefont {P.~G.}\ \bibnamefont
  {Savvidis}}, \bibinfo {author} {\bibfnamefont {J.~J.}\ \bibnamefont
  {Baumberg}}, \bibinfo {author} {\bibfnamefont {R.~M.}\ \bibnamefont
  {Stevenson}}, \bibinfo {author} {\bibfnamefont {M.~S.}\ \bibnamefont
  {Skolnick}}, \bibinfo {author} {\bibfnamefont {D.~M.}\ \bibnamefont
  {Whittaker}}, \ and\ \bibinfo {author} {\bibfnamefont {J.~S.}\ \bibnamefont
  {Roberts}},\ }\href@noop {} {\bibfield  {journal} {\bibinfo  {journal} {Phys.
  Rev. Lett.}\ }\textbf {\bibinfo {volume} {84}},\ \bibinfo {pages} {1547}
  (\bibinfo {year} {2000})}\BibitemShut {NoStop}%
\bibitem [{\citenamefont {Stevenson}\ \emph {et~al.}(2000)\citenamefont
  {Stevenson}, \citenamefont {Astratov}, \citenamefont {Skolnick},
  \citenamefont {Whittaker}, \citenamefont {Emam-Ismail}, \citenamefont
  {Tartakovskii}, \citenamefont {Savvidis}, \citenamefont {Baumberg},\ and\
  \citenamefont {Roberts}}]{stevenson00a}%
  \BibitemOpen
  \bibfield  {author} {\bibinfo {author} {\bibfnamefont {R.~M.}\ \bibnamefont
  {Stevenson}}, \bibinfo {author} {\bibfnamefont {V.~N.}\ \bibnamefont
  {Astratov}}, \bibinfo {author} {\bibfnamefont {M.~S.}\ \bibnamefont
  {Skolnick}}, \bibinfo {author} {\bibfnamefont {D.~M.}\ \bibnamefont
  {Whittaker}}, \bibinfo {author} {\bibfnamefont {M.}~\bibnamefont
  {Emam-Ismail}}, \bibinfo {author} {\bibfnamefont {A.~I.}\ \bibnamefont
  {Tartakovskii}}, \bibinfo {author} {\bibfnamefont {P.~G.}\ \bibnamefont
  {Savvidis}}, \bibinfo {author} {\bibfnamefont {J.~J.}\ \bibnamefont
  {Baumberg}}, \ and\ \bibinfo {author} {\bibfnamefont {J.~S.}\ \bibnamefont
  {Roberts}},\ }\href@noop {} {\bibfield  {journal} {\bibinfo  {journal} {Phys.
  Rev. Lett.}\ }\textbf {\bibinfo {volume} {85}},\ \bibinfo {pages} {3680}
  (\bibinfo {year} {2000})}\BibitemShut {NoStop}%
\bibitem [{\citenamefont {Ciuti}\ \emph {et~al.}(2001)\citenamefont {Ciuti},
  \citenamefont {Schwendimann},\ and\ \citenamefont {Quattropani}}]{ciuti01a}%
  \BibitemOpen
  \bibfield  {author} {\bibinfo {author} {\bibfnamefont {C.}~\bibnamefont
  {Ciuti}}, \bibinfo {author} {\bibfnamefont {P.}~\bibnamefont {Schwendimann}},
  \ and\ \bibinfo {author} {\bibfnamefont {A.}~\bibnamefont {Quattropani}},\
  }\href@noop {} {\bibfield  {journal} {\bibinfo  {journal} {Phys. Rev. B}\
  }\textbf {\bibinfo {volume} {63}},\ \bibinfo {pages} {041303} (\bibinfo
  {year} {2001})}\BibitemShut {NoStop}%
\bibitem [{\citenamefont {Ciuti}\ and\ \citenamefont
  {Carusotto}(2005)}]{ciuti05a}%
  \BibitemOpen
  \bibfield  {author} {\bibinfo {author} {\bibfnamefont {C.}~\bibnamefont
  {Ciuti}}\ and\ \bibinfo {author} {\bibfnamefont {I.}~\bibnamefont
  {Carusotto}},\ }\href@noop {} {\bibfield  {journal} {\bibinfo  {journal}
  {Phys. Stat. Sol. B}\ }\textbf {\bibinfo {volume} {242}},\ \bibinfo {pages}
  {2224} (\bibinfo {year} {2005})}\BibitemShut {NoStop}%
\bibitem [{\citenamefont {Shelykh}\ \emph {et~al.}(2006)\citenamefont
  {Shelykh}, \citenamefont {Rubo}, \citenamefont {Malpuech}, \citenamefont
  {Solnyshkov},\ and\ \citenamefont {Kavokin}}]{shelykh06a}%
  \BibitemOpen
  \bibfield  {author} {\bibinfo {author} {\bibfnamefont {I.~A.}\ \bibnamefont
  {Shelykh}}, \bibinfo {author} {\bibfnamefont {Y.~G.}\ \bibnamefont {Rubo}},
  \bibinfo {author} {\bibfnamefont {G.}~\bibnamefont {Malpuech}}, \bibinfo
  {author} {\bibfnamefont {D.~D.}\ \bibnamefont {Solnyshkov}}, \ and\ \bibinfo
  {author} {\bibfnamefont {A.}~\bibnamefont {Kavokin}},\ }\href@noop {}
  {\bibfield  {journal} {\bibinfo  {journal} {Phys. Rev. Lett.}\ }\textbf
  {\bibinfo {volume} {97}},\ \bibinfo {pages} {066402} (\bibinfo {year}
  {2006})}\BibitemShut {NoStop}%
\bibitem [{\citenamefont {Kavokin}\ \emph {et~al.}(2003)\citenamefont
  {Kavokin}, \citenamefont {Malpuech},\ and\ \citenamefont
  {Laussy}}]{kavokin03a}%
  \BibitemOpen
  \bibfield  {author} {\bibinfo {author} {\bibfnamefont {A.}~\bibnamefont
  {Kavokin}}, \bibinfo {author} {\bibfnamefont {G.}~\bibnamefont {Malpuech}}, \
  and\ \bibinfo {author} {\bibfnamefont {F.~P.}\ \bibnamefont {Laussy}},\
  }\href@noop {} {\bibfield  {journal} {\bibinfo  {journal} {Phys. Lett. A}\
  }\textbf {\bibinfo {volume} {306}},\ \bibinfo {pages} {187} (\bibinfo {year}
  {2003})}\BibitemShut {NoStop}%
\bibitem [{\citenamefont {Laussy}\ \emph {et~al.}(2004)\citenamefont {Laussy},
  \citenamefont {Malpuech}, \citenamefont {Kavokin},\ and\ \citenamefont
  {Bigenwald}}]{laussy04c}%
  \BibitemOpen
  \bibfield  {author} {\bibinfo {author} {\bibfnamefont {F.~P.}\ \bibnamefont
  {Laussy}}, \bibinfo {author} {\bibfnamefont {G.}~\bibnamefont {Malpuech}},
  \bibinfo {author} {\bibfnamefont {A.}~\bibnamefont {Kavokin}}, \ and\
  \bibinfo {author} {\bibfnamefont {P.}~\bibnamefont {Bigenwald}},\ }\href@noop
  {} {\bibfield  {journal} {\bibinfo  {journal} {Phys. Rev. Lett.}\ }\textbf
  {\bibinfo {volume} {93}},\ \bibinfo {pages} {016402} (\bibinfo {year}
  {2004})}\BibitemShut {NoStop}%
\bibitem [{\citenamefont {Szymanska}\ \emph {et~al.}(2006)\citenamefont
  {Szymanska}, \citenamefont {Keeling},\ and\ \citenamefont
  {Littlewood}}]{szymanska06a}%
  \BibitemOpen
  \bibfield  {author} {\bibinfo {author} {\bibfnamefont {M.~H.}\ \bibnamefont
  {Szymanska}}, \bibinfo {author} {\bibfnamefont {J.}~\bibnamefont {Keeling}},
  \ and\ \bibinfo {author} {\bibfnamefont {P.~B.}\ \bibnamefont {Littlewood}},\
  }\href@noop {} {\bibfield  {journal} {\bibinfo  {journal} {Phys. Rev. Lett.}\
  }\textbf {\bibinfo {volume} {96}},\ \bibinfo {pages} {230602} (\bibinfo
  {year} {2006})}\BibitemShut {NoStop}%
\bibitem [{\citenamefont {Kasprzak}\ \emph {et~al.}(2006)\citenamefont
  {Kasprzak}, \citenamefont {Richard}, \citenamefont {Kundermann},
  \citenamefont {Baas}, \citenamefont {Jeambrun}, \citenamefont {Keeling},
  \citenamefont {Marchetti}, \citenamefont {Szymanska}, \citenamefont
  {Andr\'e}, \citenamefont {Staehli}, \citenamefont {Savona}, \citenamefont
  {Littlewood}, \citenamefont {Deveaud},\ and\ \citenamefont {{Le Si
  Dang}}}]{kasprzak06a}%
  \BibitemOpen
  \bibfield  {author} {\bibinfo {author} {\bibfnamefont {J.}~\bibnamefont
  {Kasprzak}}, \bibinfo {author} {\bibfnamefont {M.}~\bibnamefont {Richard}},
  \bibinfo {author} {\bibfnamefont {S.}~\bibnamefont {Kundermann}}, \bibinfo
  {author} {\bibfnamefont {A.}~\bibnamefont {Baas}}, \bibinfo {author}
  {\bibfnamefont {P.}~\bibnamefont {Jeambrun}}, \bibinfo {author}
  {\bibfnamefont {J.~M.~J.}\ \bibnamefont {Keeling}}, \bibinfo {author}
  {\bibfnamefont {F.~M.}\ \bibnamefont {Marchetti}}, \bibinfo {author}
  {\bibfnamefont {M.~H.}\ \bibnamefont {Szymanska}}, \bibinfo {author}
  {\bibfnamefont {R.}~\bibnamefont {Andr\'e}}, \bibinfo {author} {\bibfnamefont
  {J.~L.}\ \bibnamefont {Staehli}}, \bibinfo {author} {\bibfnamefont
  {V.}~\bibnamefont {Savona}}, \bibinfo {author} {\bibfnamefont {P.~B.}\
  \bibnamefont {Littlewood}}, \bibinfo {author} {\bibfnamefont
  {B.}~\bibnamefont {Deveaud}}, \ and\ \bibinfo {author} {\bibnamefont {{Le Si
  Dang}}},\ }\href@noop {} {\bibfield  {journal} {\bibinfo  {journal} {Nature}\
  }\textbf {\bibinfo {volume} {443}},\ \bibinfo {pages} {409} (\bibinfo {year}
  {2006})}\BibitemShut {NoStop}%
\bibitem [{\citenamefont {Balili}\ \emph {et~al.}(2007)\citenamefont {Balili},
  \citenamefont {Hartwell}, \citenamefont {Snoke}, \citenamefont {Pfeiffer},\
  and\ \citenamefont {West}}]{balili07a}%
  \BibitemOpen
  \bibfield  {author} {\bibinfo {author} {\bibfnamefont {R.}~\bibnamefont
  {Balili}}, \bibinfo {author} {\bibfnamefont {V.}~\bibnamefont {Hartwell}},
  \bibinfo {author} {\bibfnamefont {D.}~\bibnamefont {Snoke}}, \bibinfo
  {author} {\bibfnamefont {L.}~\bibnamefont {Pfeiffer}}, \ and\ \bibinfo
  {author} {\bibfnamefont {K.}~\bibnamefont {West}},\ }\href@noop {} {\bibfield
   {journal} {\bibinfo  {journal} {Science}\ }\textbf {\bibinfo {volume}
  {316}},\ \bibinfo {pages} {1007} (\bibinfo {year} {2007})}\BibitemShut
  {NoStop}%
\bibitem [{\citenamefont {del Valle}\ \emph {et~al.}(2009)\citenamefont {del
  Valle}, \citenamefont {Sanvitto}, \citenamefont {Amo}, \citenamefont
  {Laussy}, \citenamefont {Andr\'e}, \citenamefont {Tejedor},\ and\
  \citenamefont {Vi{\~n}a}}]{delvalle09c}%
  \BibitemOpen
  \bibfield  {author} {\bibinfo {author} {\bibfnamefont {E.}~\bibnamefont {del
  Valle}}, \bibinfo {author} {\bibfnamefont {D.}~\bibnamefont {Sanvitto}},
  \bibinfo {author} {\bibfnamefont {A.}~\bibnamefont {Amo}}, \bibinfo {author}
  {\bibfnamefont {F.~P.}\ \bibnamefont {Laussy}}, \bibinfo {author}
  {\bibfnamefont {R.}~\bibnamefont {Andr\'e}}, \bibinfo {author} {\bibfnamefont
  {C.}~\bibnamefont {Tejedor}}, \ and\ \bibinfo {author} {\bibfnamefont
  {L.}~\bibnamefont {Vi{\~n}a}},\ }\href@noop {} {\bibfield  {journal}
  {\bibinfo  {journal} {Phys. Rev. Lett.}\ }\textbf {\bibinfo {volume} {103}},\
  \bibinfo {pages} {096404} (\bibinfo {year} {2009})}\BibitemShut {NoStop}%
\bibitem [{\citenamefont {Christopoulos}\ \emph {et~al.}(2007)\citenamefont
  {Christopoulos}, \citenamefont {von H\"ogersthal}, \citenamefont {Grundy},
  \citenamefont {Lagoudakis}, \citenamefont {Kavokin}, \citenamefont
  {Baumberg}, \citenamefont {Christmann}, \citenamefont {Butt\'e},
  \citenamefont {Feltin}, \citenamefont {Carlin},\ and\ \citenamefont
  {Grandjean}}]{christopoulos07a}%
  \BibitemOpen
  \bibfield  {author} {\bibinfo {author} {\bibfnamefont {S.}~\bibnamefont
  {Christopoulos}}, \bibinfo {author} {\bibfnamefont {G.~B.~H.}\ \bibnamefont
  {von H\"ogersthal}}, \bibinfo {author} {\bibfnamefont {A.~J.~D.}\
  \bibnamefont {Grundy}}, \bibinfo {author} {\bibfnamefont {P.~G.}\
  \bibnamefont {Lagoudakis}}, \bibinfo {author} {\bibfnamefont {A.~V.}\
  \bibnamefont {Kavokin}}, \bibinfo {author} {\bibfnamefont {J.~J.}\
  \bibnamefont {Baumberg}}, \bibinfo {author} {\bibfnamefont {G.}~\bibnamefont
  {Christmann}}, \bibinfo {author} {\bibfnamefont {R.}~\bibnamefont {Butt\'e}},
  \bibinfo {author} {\bibfnamefont {E.}~\bibnamefont {Feltin}}, \bibinfo
  {author} {\bibfnamefont {J.-F.}\ \bibnamefont {Carlin}}, \ and\ \bibinfo
  {author} {\bibfnamefont {N.}~\bibnamefont {Grandjean}},\ }\href@noop {}
  {\bibfield  {journal} {\bibinfo  {journal} {Phys. Rev. Lett.}\ }\textbf
  {\bibinfo {volume} {98}},\ \bibinfo {pages} {126405} (\bibinfo {year}
  {2007})}\BibitemShut {NoStop}%
\bibitem [{\citenamefont {Baumberg}\ \emph {et~al.}(2008)\citenamefont
  {Baumberg}, \citenamefont {Kavokin}, \citenamefont {Christopoulos},
  \citenamefont {Grundy}, \citenamefont {Butt\'e}, \citenamefont {Christmann},
  \citenamefont {Solnyshkov}, \citenamefont {Malpuech}, \citenamefont {von
  H\"ogersthal}, \citenamefont {Feltin}, \citenamefont {Carlin},\ and\
  \citenamefont {Grandjean}}]{baumberg08a}%
  \BibitemOpen
  \bibfield  {author} {\bibinfo {author} {\bibfnamefont {J.~J.}\ \bibnamefont
  {Baumberg}}, \bibinfo {author} {\bibfnamefont {A.~V.}\ \bibnamefont
  {Kavokin}}, \bibinfo {author} {\bibfnamefont {S.}~\bibnamefont
  {Christopoulos}}, \bibinfo {author} {\bibfnamefont {A.~J.~D.}\ \bibnamefont
  {Grundy}}, \bibinfo {author} {\bibfnamefont {R.}~\bibnamefont {Butt\'e}},
  \bibinfo {author} {\bibfnamefont {G.}~\bibnamefont {Christmann}}, \bibinfo
  {author} {\bibfnamefont {D.~D.}\ \bibnamefont {Solnyshkov}}, \bibinfo
  {author} {\bibfnamefont {G.}~\bibnamefont {Malpuech}}, \bibinfo {author}
  {\bibfnamefont {G.~B.~H.}\ \bibnamefont {von H\"ogersthal}}, \bibinfo
  {author} {\bibfnamefont {E.}~\bibnamefont {Feltin}}, \bibinfo {author}
  {\bibfnamefont {J.-F.}\ \bibnamefont {Carlin}}, \ and\ \bibinfo {author}
  {\bibfnamefont {N.}~\bibnamefont {Grandjean}},\ }\href@noop {} {\bibfield
  {journal} {\bibinfo  {journal} {Phys. Rev. Lett.}\ }\textbf {\bibinfo
  {volume} {101}},\ \bibinfo {pages} {136409} (\bibinfo {year}
  {2008})}\BibitemShut {NoStop}%
\bibitem [{\citenamefont {Amo}\ \emph {et~al.}(2009{\natexlab{a}})\citenamefont
  {Amo}, \citenamefont {Sanvitto}, \citenamefont {Laussy}, \citenamefont
  {Ballarini}, \citenamefont {del Valle}, \citenamefont {Martin}, \citenamefont
  {Lema\^itre}, \citenamefont {Bloch}, \citenamefont {Krizhanovskii},
  \citenamefont {Skolnick}, \citenamefont {Tejedor},\ and\ \citenamefont
  {Vi{\~n}a}}]{amo09a}%
  \BibitemOpen
  \bibfield  {author} {\bibinfo {author} {\bibfnamefont {A.}~\bibnamefont
  {Amo}}, \bibinfo {author} {\bibfnamefont {D.}~\bibnamefont {Sanvitto}},
  \bibinfo {author} {\bibfnamefont {F.~P.}\ \bibnamefont {Laussy}}, \bibinfo
  {author} {\bibfnamefont {D.}~\bibnamefont {Ballarini}}, \bibinfo {author}
  {\bibfnamefont {E.}~\bibnamefont {del Valle}}, \bibinfo {author}
  {\bibfnamefont {M.~D.}\ \bibnamefont {Martin}}, \bibinfo {author}
  {\bibfnamefont {A.}~\bibnamefont {Lema\^itre}}, \bibinfo {author}
  {\bibfnamefont {J.}~\bibnamefont {Bloch}}, \bibinfo {author} {\bibfnamefont
  {D.~N.}\ \bibnamefont {Krizhanovskii}}, \bibinfo {author} {\bibfnamefont
  {M.~S.}\ \bibnamefont {Skolnick}}, \bibinfo {author} {\bibfnamefont
  {C.}~\bibnamefont {Tejedor}}, \ and\ \bibinfo {author} {\bibfnamefont
  {L.}~\bibnamefont {Vi{\~n}a}},\ }\href@noop {} {\bibfield  {journal}
  {\bibinfo  {journal} {Nature}\ }\textbf {\bibinfo {volume} {457}},\ \bibinfo
  {pages} {291} (\bibinfo {year} {2009}{\natexlab{a}})}\BibitemShut {NoStop}%
\bibitem [{\citenamefont {Sanvitto}\ \emph
  {et~al.}(2010{\natexlab{a}})\citenamefont {Sanvitto}, \citenamefont {Amo},
  \citenamefont {Laussy}, \citenamefont {Lema{\^i}tre}, \citenamefont {Bloch},
  \citenamefont {Tejedor},\ and\ \citenamefont {Vi{\~n}a}}]{sanvitto10a}%
  \BibitemOpen
  \bibfield  {author} {\bibinfo {author} {\bibfnamefont {D.}~\bibnamefont
  {Sanvitto}}, \bibinfo {author} {\bibfnamefont {A.}~\bibnamefont {Amo}},
  \bibinfo {author} {\bibfnamefont {F.~P.}\ \bibnamefont {Laussy}}, \bibinfo
  {author} {\bibfnamefont {A.}~\bibnamefont {Lema{\^i}tre}}, \bibinfo {author}
  {\bibfnamefont {J.}~\bibnamefont {Bloch}}, \bibinfo {author} {\bibfnamefont
  {C.}~\bibnamefont {Tejedor}}, \ and\ \bibinfo {author} {\bibfnamefont
  {L.}~\bibnamefont {Vi{\~n}a}},\ }\href@noop {} {\bibfield  {journal}
  {\bibinfo  {journal} {Nanotechnology}\ }\textbf {\bibinfo {volume} {21}},\
  \bibinfo {pages} {134025} (\bibinfo {year} {2010}{\natexlab{a}})}\BibitemShut
  {NoStop}%
\bibitem [{\citenamefont {Amo}\ \emph {et~al.}(2009{\natexlab{b}})\citenamefont
  {Amo}, \citenamefont {Lefr\`ere}, \citenamefont {Pigeon}, \citenamefont
  {Adrados}, \citenamefont {Ciuti}, \citenamefont {Carusotto}, \citenamefont
  {Houdr\'e}, \citenamefont {Giacobino},\ and\ \citenamefont
  {Bramati}}]{amo09b}%
  \BibitemOpen
  \bibfield  {author} {\bibinfo {author} {\bibfnamefont {A.}~\bibnamefont
  {Amo}}, \bibinfo {author} {\bibfnamefont {J.}~\bibnamefont {Lefr\`ere}},
  \bibinfo {author} {\bibfnamefont {S.}~\bibnamefont {Pigeon}}, \bibinfo
  {author} {\bibfnamefont {C.}~\bibnamefont {Adrados}}, \bibinfo {author}
  {\bibfnamefont {C.}~\bibnamefont {Ciuti}}, \bibinfo {author} {\bibfnamefont
  {I.}~\bibnamefont {Carusotto}}, \bibinfo {author} {\bibfnamefont
  {R.}~\bibnamefont {Houdr\'e}}, \bibinfo {author} {\bibfnamefont
  {E.}~\bibnamefont {Giacobino}}, \ and\ \bibinfo {author} {\bibfnamefont
  {A.}~\bibnamefont {Bramati}},\ }\href@noop {} {\bibfield  {journal} {\bibinfo
   {journal} {Nat. Phys.}\ }\textbf {\bibinfo {volume} {5}},\ \bibinfo {pages}
  {805} (\bibinfo {year} {2009}{\natexlab{b}})}\BibitemShut {NoStop}%
\bibitem [{\citenamefont {Lagoudakis}\ \emph {et~al.}(2008)\citenamefont
  {Lagoudakis}, \citenamefont {Wouters}, \citenamefont {Richard}, \citenamefont
  {Baas}, \citenamefont {Carusotto}, \citenamefont {Andr{\'e}}, \citenamefont
  {Dang},\ and\ \citenamefont {Deveaud-Pl{\'e}dran}}]{lagoudakis08a}%
  \BibitemOpen
  \bibfield  {author} {\bibinfo {author} {\bibfnamefont {K.~G.}\ \bibnamefont
  {Lagoudakis}}, \bibinfo {author} {\bibfnamefont {M.}~\bibnamefont {Wouters}},
  \bibinfo {author} {\bibfnamefont {M.}~\bibnamefont {Richard}}, \bibinfo
  {author} {\bibfnamefont {A.}~\bibnamefont {Baas}}, \bibinfo {author}
  {\bibfnamefont {I.}~\bibnamefont {Carusotto}}, \bibinfo {author}
  {\bibfnamefont {R.}~\bibnamefont {Andr{\'e}}}, \bibinfo {author}
  {\bibfnamefont {L.~S.}\ \bibnamefont {Dang}}, \ and\ \bibinfo {author}
  {\bibfnamefont {B.}~\bibnamefont {Deveaud-Pl{\'e}dran}},\ }\href@noop {}
  {\bibfield  {journal} {\bibinfo  {journal} {Nat. Phys.}\ }\textbf {\bibinfo
  {volume} {4}},\ \bibinfo {pages} {706} (\bibinfo {year} {2008})}\BibitemShut
  {NoStop}%
\bibitem [{\citenamefont {Lagoudakis}\ \emph {et~al.}(2009)\citenamefont
  {Lagoudakis}, \citenamefont {Ostatnick\'y}, \citenamefont {Kavokin},
  \citenamefont {Rubo}, \citenamefont {Andr\'e},\ and\ \citenamefont
  {Deveaud-Pl\'edran}}]{lagoudakis09a}%
  \BibitemOpen
  \bibfield  {author} {\bibinfo {author} {\bibfnamefont {K.~G.}\ \bibnamefont
  {Lagoudakis}}, \bibinfo {author} {\bibfnamefont {T.}~\bibnamefont
  {Ostatnick\'y}}, \bibinfo {author} {\bibfnamefont {A.~V.}\ \bibnamefont
  {Kavokin}}, \bibinfo {author} {\bibfnamefont {Y.~G.}\ \bibnamefont {Rubo}},
  \bibinfo {author} {\bibfnamefont {R.}~\bibnamefont {Andr\'e}}, \ and\
  \bibinfo {author} {\bibfnamefont {B.}~\bibnamefont {Deveaud-Pl\'edran}},\
  }\href@noop {} {\bibfield  {journal} {\bibinfo  {journal} {Science}\ }\textbf
  {\bibinfo {volume} {326}},\ \bibinfo {pages} {974} (\bibinfo {year}
  {2009})}\BibitemShut {NoStop}%
\bibitem [{\citenamefont {Sanvitto}\ \emph
  {et~al.}(2010{\natexlab{b}})\citenamefont {Sanvitto}, \citenamefont
  {Marchetti}, \citenamefont {Szyma{\'n}ska}, \citenamefont {Tosi},
  \citenamefont {Baudisch}, \citenamefont {Laussy}, \citenamefont
  {Krizhanovskii}, \citenamefont {Skolnick}, \citenamefont {Marrucci},
  \citenamefont {Lema{\^i}tre}, \citenamefont {Bloch}, \citenamefont
  {Tejedor},\ and\ \citenamefont {Vi{\~n}a}}]{sanvitto10b}%
  \BibitemOpen
  \bibfield  {author} {\bibinfo {author} {\bibfnamefont {D.}~\bibnamefont
  {Sanvitto}}, \bibinfo {author} {\bibfnamefont {F.~M.}\ \bibnamefont
  {Marchetti}}, \bibinfo {author} {\bibfnamefont {M.~H.}\ \bibnamefont
  {Szyma{\'n}ska}}, \bibinfo {author} {\bibfnamefont {G.}~\bibnamefont {Tosi}},
  \bibinfo {author} {\bibfnamefont {M.}~\bibnamefont {Baudisch}}, \bibinfo
  {author} {\bibfnamefont {F.~P.}\ \bibnamefont {Laussy}}, \bibinfo {author}
  {\bibfnamefont {D.~N.}\ \bibnamefont {Krizhanovskii}}, \bibinfo {author}
  {\bibfnamefont {M.~S.}\ \bibnamefont {Skolnick}}, \bibinfo {author}
  {\bibfnamefont {L.}~\bibnamefont {Marrucci}}, \bibinfo {author}
  {\bibfnamefont {A.}~\bibnamefont {Lema{\^i}tre}}, \bibinfo {author}
  {\bibfnamefont {J.}~\bibnamefont {Bloch}}, \bibinfo {author} {\bibfnamefont
  {C.}~\bibnamefont {Tejedor}}, \ and\ \bibinfo {author} {\bibfnamefont
  {L.}~\bibnamefont {Vi{\~n}a}},\ }\href@noop {} {\bibfield  {journal}
  {\bibinfo  {journal} {Nat. Phys.}\ }\textbf {\bibinfo {volume} {6}},\
  \bibinfo {pages} {527} (\bibinfo {year} {2010}{\natexlab{b}})}\BibitemShut
  {NoStop}%
\bibitem [{\citenamefont {Kavokin}\ \emph
  {et~al.}(2007{\natexlab{b}})\citenamefont {Kavokin}, \citenamefont
  {Solnyshkov},\ and\ \citenamefont {Malpuech}}]{kavokin07b}%
  \BibitemOpen
  \bibfield  {author} {\bibinfo {author} {\bibfnamefont {A.}~\bibnamefont
  {Kavokin}}, \bibinfo {author} {\bibfnamefont {D.}~\bibnamefont {Solnyshkov}},
  \ and\ \bibinfo {author} {\bibfnamefont {G.}~\bibnamefont {Malpuech}},\
  }\href@noop {} {\bibfield  {journal} {\bibinfo  {journal} {J. Phys.: Condens.
  Matter}\ }\textbf {\bibinfo {volume} {19}},\ \bibinfo {pages} {295212}
  (\bibinfo {year} {2007}{\natexlab{b}})}\BibitemShut {NoStop}%
\bibitem [{\citenamefont {Laussy}\ \emph {et~al.}(2010)\citenamefont {Laussy},
  \citenamefont {Kavokin},\ and\ \citenamefont {Shelykh}}]{laussy10a}%
  \BibitemOpen
  \bibfield  {author} {\bibinfo {author} {\bibfnamefont {F.~P.}\ \bibnamefont
  {Laussy}}, \bibinfo {author} {\bibfnamefont {A.~V.}\ \bibnamefont {Kavokin}},
  \ and\ \bibinfo {author} {\bibfnamefont {I.~A.}\ \bibnamefont {Shelykh}},\
  }\href@noop {} {\bibfield  {journal} {\bibinfo  {journal} {Phys. Rev. Lett.}\
  }\textbf {\bibinfo {volume} {104}},\ \bibinfo {pages} {106402} (\bibinfo
  {year} {2010})}\BibitemShut {NoStop}%
\bibitem [{\citenamefont {Cooper}(1956)}]{cooper56a}%
  \BibitemOpen
  \bibfield  {author} {\bibinfo {author} {\bibfnamefont {L.~N.}\ \bibnamefont
  {Cooper}},\ }\href@noop {} {\bibfield  {journal} {\bibinfo  {journal} {Phys.
  Rev.}\ }\textbf {\bibinfo {volume} {104}},\ \bibinfo {pages} {1189} (\bibinfo
  {year} {1956})}\BibitemShut {NoStop}%
\bibitem [{\citenamefont {Bardeen}\ and\ \citenamefont
  {Pines}(1955)}]{bardeen55b}%
  \BibitemOpen
  \bibfield  {author} {\bibinfo {author} {\bibfnamefont {J.}~\bibnamefont
  {Bardeen}}\ and\ \bibinfo {author} {\bibfnamefont {D.}~\bibnamefont
  {Pines}},\ }\href@noop {} {\bibfield  {journal} {\bibinfo  {journal} {Phys.
  Rev.}\ }\textbf {\bibinfo {volume} {99}},\ \bibinfo {pages} {1140} (\bibinfo
  {year} {1955})}\BibitemShut {NoStop}%
\bibitem [{\citenamefont {Bardeen}(1951)}]{bardeen51a}%
  \BibitemOpen
  \bibfield  {author} {\bibinfo {author} {\bibfnamefont {J.}~\bibnamefont
  {Bardeen}},\ }\href@noop {} {\bibfield  {journal} {\bibinfo  {journal} {Rev.
  Mod. Phys.}\ }\textbf {\bibinfo {volume} {23}},\ \bibinfo {pages} {261}
  (\bibinfo {year} {1951})}\BibitemShut {NoStop}%
\bibitem [{\citenamefont {Leggett}(2006{\natexlab{a}})}]{leggett_book06a}%
  \BibitemOpen
  \bibfield  {author} {\bibinfo {author} {\bibfnamefont {A.~J.}\ \bibnamefont
  {Leggett}},\ }\href@noop {} {\emph {\bibinfo {title} {Quantum Liquids}}}\
  (\bibinfo  {publisher} {Oxford University Press},\ \bibinfo {year}
  {2006})\BibitemShut {NoStop}%
\bibitem [{\citenamefont {Bardeen}\ \emph
  {et~al.}(1957{\natexlab{a}})\citenamefont {Bardeen}, \citenamefont {Cooper},\
  and\ \citenamefont {Schrieffer}}]{bardeen57b}%
  \BibitemOpen
  \bibfield  {author} {\bibinfo {author} {\bibfnamefont {J.}~\bibnamefont
  {Bardeen}}, \bibinfo {author} {\bibfnamefont {L.~N.}\ \bibnamefont {Cooper}},
  \ and\ \bibinfo {author} {\bibfnamefont {J.~R.}\ \bibnamefont {Schrieffer}},\
  }\href@noop {} {\bibfield  {journal} {\bibinfo  {journal} {Phys. Rev.}\
  }\textbf {\bibinfo {volume} {106}},\ \bibinfo {pages} {162} (\bibinfo {year}
  {1957}{\natexlab{a}})}\BibitemShut {NoStop}%
\bibitem [{\citenamefont {Bardeen}\ \emph
  {et~al.}(1957{\natexlab{b}})\citenamefont {Bardeen}, \citenamefont {Cooper},\
  and\ \citenamefont {Schrieffer}}]{bardeen57a}%
  \BibitemOpen
  \bibfield  {author} {\bibinfo {author} {\bibfnamefont {J.}~\bibnamefont
  {Bardeen}}, \bibinfo {author} {\bibfnamefont {L.~N.}\ \bibnamefont {Cooper}},
  \ and\ \bibinfo {author} {\bibfnamefont {J.~R.}\ \bibnamefont {Schrieffer}},\
  }\href@noop {} {\bibfield  {journal} {\bibinfo  {journal} {Phys. Rev.}\
  }\textbf {\bibinfo {volume} {108}},\ \bibinfo {pages} {1175} (\bibinfo {year}
  {1957}{\natexlab{b}})}\BibitemShut {NoStop}%
\bibitem [{\citenamefont {Allender}\ \emph {et~al.}(1973)\citenamefont
  {Allender}, \citenamefont {Bray},\ and\ \citenamefont
  {Bardeen}}]{allender73a}%
  \BibitemOpen
  \bibfield  {author} {\bibinfo {author} {\bibfnamefont {D.}~\bibnamefont
  {Allender}}, \bibinfo {author} {\bibfnamefont {J.}~\bibnamefont {Bray}}, \
  and\ \bibinfo {author} {\bibfnamefont {J.}~\bibnamefont {Bardeen}},\
  }\href@noop {} {\bibfield  {journal} {\bibinfo  {journal} {Phys. Rev. B}\
  }\textbf {\bibinfo {volume} {7}},\ \bibinfo {pages} {1020} (\bibinfo {year}
  {1973})}\BibitemShut {NoStop}%
\bibitem [{\citenamefont {Little}(1964)}]{little64a}%
  \BibitemOpen
  \bibfield  {author} {\bibinfo {author} {\bibfnamefont {W.~A.}\ \bibnamefont
  {Little}},\ }\href@noop {} {\bibfield  {journal} {\bibinfo  {journal} {Phys.
  Rev.}\ }\textbf {\bibinfo {volume} {134}},\ \bibinfo {pages} {A1416}
  (\bibinfo {year} {1964})}\BibitemShut {NoStop}%
\bibitem [{\citenamefont {Ginzburg}(1970)}]{sp_ginzburg70a}%
  \BibitemOpen
  \bibfield  {author} {\bibinfo {author} {\bibfnamefont {V.~L.}\ \bibnamefont
  {Ginzburg}},\ }\href@noop {} {\bibfield  {journal} {\bibinfo  {journal} {Sov.
  Phys.-Usp.}\ }\textbf {\bibinfo {volume} {13}},\ \bibinfo {pages} {335}
  (\bibinfo {year} {1970})}\BibitemShut {NoStop}%
\bibitem [{\citenamefont {Bednorz}\ and\ \citenamefont
  {M\"uller}(1986)}]{bednorz86a}%
  \BibitemOpen
  \bibfield  {author} {\bibinfo {author} {\bibfnamefont {J.~G.}\ \bibnamefont
  {Bednorz}}\ and\ \bibinfo {author} {\bibfnamefont {K.~A.}\ \bibnamefont
  {M\"uller}},\ }\href@noop {} {\bibfield  {journal} {\bibinfo  {journal} {Z.
  Phys. B}\ }\textbf {\bibinfo {volume} {64}},\ \bibinfo {pages} {189}
  (\bibinfo {year} {1986})}\BibitemShut {NoStop}%
\bibitem [{\citenamefont {Leggett}(2006{\natexlab{b}})}]{leggett06a}%
  \BibitemOpen
  \bibfield  {author} {\bibinfo {author} {\bibfnamefont {A.~J.}\ \bibnamefont
  {Leggett}},\ }\href@noop {} {\bibfield  {journal} {\bibinfo  {journal} {Nat.
  Phys.}\ }\textbf {\bibinfo {volume} {2}},\ \bibinfo {pages} {134} (\bibinfo
  {year} {2006}{\natexlab{b}})}\BibitemShut {NoStop}%
\bibitem [{\citenamefont {Butov}\ \emph {et~al.}(2002)\citenamefont {Butov},
  \citenamefont {Lai}, \citenamefont {Ivanov}, \citenamefont {Gossard},\ and\
  \citenamefont {Chemla}}]{butov02a}%
  \BibitemOpen
  \bibfield  {author} {\bibinfo {author} {\bibfnamefont {L.~V.}\ \bibnamefont
  {Butov}}, \bibinfo {author} {\bibfnamefont {C.~W.}\ \bibnamefont {Lai}},
  \bibinfo {author} {\bibfnamefont {A.~L.}\ \bibnamefont {Ivanov}}, \bibinfo
  {author} {\bibfnamefont {A.~C.}\ \bibnamefont {Gossard}}, \ and\ \bibinfo
  {author} {\bibfnamefont {D.~S.}\ \bibnamefont {Chemla}},\ }\href@noop {}
  {\bibfield  {journal} {\bibinfo  {journal} {Nature}\ }\textbf {\bibinfo
  {volume} {417}},\ \bibinfo {pages} {47} (\bibinfo {year} {2002})}\BibitemShut
  {NoStop}%
\bibitem [{\citenamefont {Butov}\ \emph {et~al.}(2004)\citenamefont {Butov},
  \citenamefont {Levitov}, \citenamefont {Mintsev}, \citenamefont {Simons},
  \citenamefont {Gossard},\ and\ \citenamefont {Chemla}}]{butov04a}%
  \BibitemOpen
  \bibfield  {author} {\bibinfo {author} {\bibfnamefont {L.~V.}\ \bibnamefont
  {Butov}}, \bibinfo {author} {\bibfnamefont {L.~S.}\ \bibnamefont {Levitov}},
  \bibinfo {author} {\bibfnamefont {A.~V.}\ \bibnamefont {Mintsev}}, \bibinfo
  {author} {\bibfnamefont {B.~D.}\ \bibnamefont {Simons}}, \bibinfo {author}
  {\bibfnamefont {A.~C.}\ \bibnamefont {Gossard}}, \ and\ \bibinfo {author}
  {\bibfnamefont {D.~S.}\ \bibnamefont {Chemla}},\ }\href@noop {} {\bibfield
  {journal} {\bibinfo  {journal} {Phys. Rev. Lett.}\ }\textbf {\bibinfo
  {volume} {92}},\ \bibinfo {pages} {117404} (\bibinfo {year}
  {2004})}\BibitemShut {NoStop}%
\bibitem [{\citenamefont {Rapaport}\ \emph {et~al.}(2004)\citenamefont
  {Rapaport}, \citenamefont {Chen}, \citenamefont {Snoke}, \citenamefont
  {Simon}, \citenamefont {Pfeiffer}, \citenamefont {West}, \citenamefont
  {Liu},\ and\ \citenamefont {Denev}}]{rapaport04a}%
  \BibitemOpen
  \bibfield  {author} {\bibinfo {author} {\bibfnamefont {R.}~\bibnamefont
  {Rapaport}}, \bibinfo {author} {\bibfnamefont {G.}~\bibnamefont {Chen}},
  \bibinfo {author} {\bibfnamefont {D.}~\bibnamefont {Snoke}}, \bibinfo
  {author} {\bibfnamefont {S.~H.}\ \bibnamefont {Simon}}, \bibinfo {author}
  {\bibfnamefont {L.}~\bibnamefont {Pfeiffer}}, \bibinfo {author}
  {\bibfnamefont {K.}~\bibnamefont {West}}, \bibinfo {author} {\bibfnamefont
  {Y.}~\bibnamefont {Liu}}, \ and\ \bibinfo {author} {\bibfnamefont
  {S.}~\bibnamefont {Denev}},\ }\href@noop {} {\bibfield  {journal} {\bibinfo
  {journal} {Phys. Rev. Lett.}\ }\textbf {\bibinfo {volume} {92}},\ \bibinfo
  {pages} {117405} (\bibinfo {year} {2004})}\BibitemShut {NoStop}%
\bibitem [{\citenamefont {Ginzburg}\ and\ \citenamefont
  {Kirzhnits}(1982)}]{ginzburg82a}%
  \BibitemOpen
  \bibinfo {editor} {\bibfnamefont {V.}~\bibnamefont {Ginzburg}}\ and\ \bibinfo
  {editor} {\bibfnamefont {D.}~\bibnamefont {Kirzhnits}},\ eds.,\ \href@noop {}
  {\emph {\bibinfo {title} {High-Temperature Superconductivity}}}\ (\bibinfo
  {publisher} {Springer},\ \bibinfo {year} {1982})\BibitemShut {NoStop}%
\bibitem [{\citenamefont {Monthoux}\ \emph {et~al.}(2007)\citenamefont
  {Monthoux}, \citenamefont {Pines},\ and\ \citenamefont
  {Lonzarich}}]{monthoux07a}%
  \BibitemOpen
  \bibfield  {author} {\bibinfo {author} {\bibfnamefont {P.}~\bibnamefont
  {Monthoux}}, \bibinfo {author} {\bibfnamefont {D.}~\bibnamefont {Pines}}, \
  and\ \bibinfo {author} {\bibfnamefont {G.~G.}\ \bibnamefont {Lonzarich}},\
  }\href@noop {} {\bibfield  {journal} {\bibinfo  {journal} {Nature}\ }\textbf
  {\bibinfo {volume} {450}},\ \bibinfo {pages} {1177} (\bibinfo {year}
  {2007})}\BibitemShut {NoStop}%
\bibitem [{\citenamefont {Butt\'e}\ and\ \citenamefont
  {Grandjean}(2010)}]{butte10a}%
  \BibitemOpen
  \bibfield  {author} {\bibinfo {author} {\bibfnamefont {R.}~\bibnamefont
  {Butt\'e}}\ and\ \bibinfo {author} {\bibfnamefont {N.}~\bibnamefont
  {Grandjean}},\ }\href@noop {} {\bibfield  {journal} {\bibinfo  {journal}
  {Semicond. Sci. Technol.}\ }\textbf {\bibinfo {volume} {26}},\ \bibinfo
  {pages} {014030} (\bibinfo {year} {2010})}\BibitemShut {NoStop}%
\bibitem [{\citenamefont {Tassone}\ and\ \citenamefont
  {Yamamoto}(1999)}]{tassone99a}%
  \BibitemOpen
  \bibfield  {author} {\bibinfo {author} {\bibfnamefont {F.}~\bibnamefont
  {Tassone}}\ and\ \bibinfo {author} {\bibfnamefont {Y.}~\bibnamefont
  {Yamamoto}},\ }\href@noop {} {\bibfield  {journal} {\bibinfo  {journal}
  {Phys. Rev. B}\ }\textbf {\bibinfo {volume} {59}},\ \bibinfo {pages} {10830}
  (\bibinfo {year} {1999})}\BibitemShut {NoStop}%
\bibitem [{\citenamefont {Bijlsma}\ \emph {et~al.}(2000)\citenamefont
  {Bijlsma}, \citenamefont {Heringa},\ and\ \citenamefont
  {Stoof}}]{bijlsma00a}%
  \BibitemOpen
  \bibfield  {author} {\bibinfo {author} {\bibfnamefont {M.~J.}\ \bibnamefont
  {Bijlsma}}, \bibinfo {author} {\bibfnamefont {B.~A.}\ \bibnamefont
  {Heringa}}, \ and\ \bibinfo {author} {\bibfnamefont {H.~T.~C.}\ \bibnamefont
  {Stoof}},\ }\href@noop {} {\bibfield  {journal} {\bibinfo  {journal} {Phys.
  Rev. A}\ }\textbf {\bibinfo {volume} {61}},\ \bibinfo {pages} {053601}
  (\bibinfo {year} {2000})}\BibitemShut {NoStop}%
\bibitem [{\citenamefont {Ginzburg}(2009)}]{ginzburg_book09a}%
  \BibitemOpen
  \bibfield  {author} {\bibinfo {author} {\bibfnamefont {V.~L.}\ \bibnamefont
  {Ginzburg}},\ }\href@noop {} {\emph {\bibinfo {title} {On Superconductivity
  and Superfluidity}}}\ (\bibinfo  {publisher} {Springer},\ \bibinfo {year}
  {2009})\BibitemShut {NoStop}%
\bibitem [{\citenamefont {Shelykh}\ \emph {et~al.}(2010)\citenamefont
  {Shelykh}, \citenamefont {Taylor},\ and\ \citenamefont
  {Kavokin}}]{shelykh10a}%
  \BibitemOpen
  \bibfield  {author} {\bibinfo {author} {\bibfnamefont {I.~A.}\ \bibnamefont
  {Shelykh}}, \bibinfo {author} {\bibfnamefont {T.}~\bibnamefont {Taylor}}, \
  and\ \bibinfo {author} {\bibfnamefont {A.~V.}\ \bibnamefont {Kavokin}},\
  }\href@noop {} {\bibfield  {journal} {\bibinfo  {journal} {Phys. Rev. Lett.}\
  }\textbf {\bibinfo {volume} {105}},\ \bibinfo {pages} {140402} (\bibinfo
  {year} {2010})}\BibitemShut {NoStop}%
\bibitem [{\citenamefont {Thompson}(1998)}]{thompson98a}%
  \BibitemOpen
  \bibfield  {author} {\bibinfo {author} {\bibfnamefont {W.~J.}\ \bibnamefont
  {Thompson}},\ }\href@noop {} {\bibfield  {journal} {\bibinfo  {journal}
  {Computers in Physics}\ }\textbf {\bibinfo {volume} {12}},\ \bibinfo {pages}
  {94} (\bibinfo {year} {1998})}\BibitemShut {NoStop}%
\bibitem [{\citenamefont {Noble}(2000)}]{noble00a}%
  \BibitemOpen
  \bibfield  {author} {\bibinfo {author} {\bibfnamefont {J.~V.}\ \bibnamefont
  {Noble}},\ }\href@noop {} {\bibfield  {journal} {\bibinfo  {journal}
  {Computing in Science and Engineering}\ }\textbf {\bibinfo {volume} {2}},\
  \bibinfo {pages} {92} (\bibinfo {year} {2000})}\BibitemShut {NoStop}%
\bibitem [{\citenamefont {de~Gennes}(1999)}]{degennes_book99a}%
  \BibitemOpen
  \bibfield  {author} {\bibinfo {author} {\bibfnamefont {P.~G.}\ \bibnamefont
  {de~Gennes}},\ }\href@noop {} {\emph {\bibinfo {title} {Superconductivity Of
  Metals And Alloys}}}\ (\bibinfo  {publisher} {Westview Press},\ \bibinfo
  {year} {1999})\BibitemShut {NoStop}%
\bibitem [{\citenamefont {Kitamura}(1963)}]{kitamura63a}%
  \BibitemOpen
  \bibfield  {author} {\bibinfo {author} {\bibfnamefont {M.}~\bibnamefont
  {Kitamura}},\ }\href@noop {} {\bibfield  {journal} {\bibinfo  {journal}
  {Prog. of Th. Phys.}\ }\textbf {\bibinfo {volume} {30}},\ \bibinfo {pages}
  {435} (\bibinfo {year} {1963})}\BibitemShut {NoStop}%
\bibitem [{\citenamefont {Vansevenant}(1985)}]{vansevenant85a}%
  \BibitemOpen
  \bibfield  {author} {\bibinfo {author} {\bibfnamefont {A.}~\bibnamefont
  {Vansevenant}},\ }\href@noop {} {\bibfield  {journal} {\bibinfo  {journal}
  {Physica D}\ }\textbf {\bibinfo {volume} {17}},\ \bibinfo {pages} {339}
  (\bibinfo {year} {1985})}\BibitemShut {NoStop}%
\bibitem [{\citenamefont {Zabreiko}\ and\ \citenamefont
  {Povolotskii}(1967)}]{zabreiko67a}%
  \BibitemOpen
  \bibfield  {author} {\bibinfo {author} {\bibfnamefont {P.~P.}\ \bibnamefont
  {Zabreiko}}\ and\ \bibinfo {author} {\bibfnamefont {A.~I.}\ \bibnamefont
  {Povolotskii}},\ }\href@noop {} {\bibfield  {journal} {\bibinfo  {journal}
  {Ukrainian Mathematical Journal}\ }\textbf {\bibinfo {volume} {22}},\
  \bibinfo {pages} {150} (\bibinfo {year} {1967})}\BibitemShut {NoStop}%
\bibitem [{\citenamefont {Ketterson}\ and\ \citenamefont
  {Song}(1999)}]{ketterson_book99a}%
  \BibitemOpen
  \bibfield  {author} {\bibinfo {author} {\bibfnamefont {J.~B.}\ \bibnamefont
  {Ketterson}}\ and\ \bibinfo {author} {\bibfnamefont {S.~N.}\ \bibnamefont
  {Song}},\ }\href@noop {} {\emph {\bibinfo {title} {Superconductivity}}}\
  (\bibinfo  {publisher} {Cambridge University Press},\ \bibinfo {year}
  {1999})\BibitemShut {NoStop}%
\bibitem [{\citenamefont {Zheng}\ and\ \citenamefont
  {Walmsley}(2005)}]{zheng05a}%
  \BibitemOpen
  \bibfield  {author} {\bibinfo {author} {\bibfnamefont {X.~H.}\ \bibnamefont
  {Zheng}}\ and\ \bibinfo {author} {\bibfnamefont {D.~G.}\ \bibnamefont
  {Walmsley}},\ }\href@noop {} {\bibfield  {journal} {\bibinfo  {journal}
  {Phys. Rev. B}\ }\textbf {\bibinfo {volume} {71}},\ \bibinfo {pages} {134512}
  (\bibinfo {year} {2005})}\BibitemShut {NoStop}%
\end{thebibliography}%

\end{document}